\documentclass[twocolumn,trackchanges]{aastex62}

\usepackage{rotating}
\usepackage{float}
\usepackage[caption = false]{subfig}

\received{XXXX}
\revised{XXXX}
\accepted{XXXX}

\submitjournal{ApJ}
\shorttitle{Detection of a White Dwarf companion to a Blue Straggler Star in NGC 5466 with UVIT}
\shortauthors{Snehalata et al.}

\begin{document}

\title{Detection of a White Dwarf companion to a Blue Straggler Star in the outskirts of globular cluster NGC 5466 with the Ultraviolet Imaging Telescope (UVIT) }

\author{Snehalata Sahu}
\affiliation{Indian Institute of Astrophysics, Koramangala II Block, Bangalore-560034, India}
\affiliation{Pondicherry University, R.V. Nagar, Kalapet, 605014, Puducherry, India}
\email{snehalata@iiap.res.in, snehalatash30@gmail.com}

\author{Annapurni Subramaniam}
\affiliation{Indian Institute of Astrophysics, Koramangala II Block, Bangalore-560034, India}
%\email{purni@iiap.res.in}

\author{Mirko Simunovic}
\affiliation{Science Fellow,Gemini Observatory, 670 N Aohoku Pl, Hilo, Hawaii 96720}

\author{J. Postma}
\affiliation{University of Calgary, Calgary, Alberta, Canada}

\author{Patrick C\^ot\'e}
\affiliation{National Research Council of Canada, Herzberg Astronomy and Astrophysics, 5071 West Saanich road, Victoria, BC V9E 2E7, Canada}
%\email{purni@iiap.res.in}

\author{N.Kameswera Rao}
\affiliation{Indian Institute of Astrophysics, Koramangala II Block, Bangalore-560034, India}
%\email{purni@iiap.res.in}

\author{Aaron M. Geller}
\affiliation{Center for Interdisciplinary Exploration and Research in Astrophysics (CIERA) and Department of Physics \& Astronomy, Northwestern University, 2145 Sheridan Rd., Evanston, IL 60201, USA}
\affiliation{Adler Planetarium, Dept. of Astronomy, 1300 S. Lake Shore Drive, Chicago, IL 60605, USA}

\author{Nathan Leigh}
\affiliation{Department of Physics and Astronomy, Stony Brook University, Stony Brook, NY 11794-3800, USA}
\affiliation{Department of Astrophysics, American Museum of Natural History, New York, NY 10024, USA}
\affiliation{Departamento de Astronom\'ia, Facultad de Ciencias F\'isicas y Matem\'aticas, Universidad de Concepci\'on, Concepci\'on, Chile}

\author{Michael Shara}
\affiliation{Department of Astrophysics, American Museum of Natural History, Central Park West and 79th Street, New York, NY 10024}

\author[0000-0003-0350-7061]{Thomas H.~Puzia}
\affiliation{Institute of Astrophysics, Pontificia Universidad Cat\'olica de Chile, Av.~Vicu\~na Mackenna 4860, 7820436 Macul, Santiago, Chile}

\author{Peter B. Stetson}
\affiliation{National Research Council of Canada, Herzberg Astronomy and Astrophysics, 5071 West Saanich road, Victoria, BC V9E 2E7, Canada}

\begin{abstract}
We report the discovery of a hot white dwarf (WD) companion to a blue straggler star (BSS) in the globular cluster (GC) NGC 5466, based on observations from the Ultra-Violet Imaging Telescope (UVIT) on board AstroSat. The Spectral Energy Distribution (SED) of the Far-UV detected BSS NH 84 was constructed by combining the flux measurements from 4 filters of UVIT, with GALEX, GAIA and other ground-based observations. The SED of NH 84 reveals the presence of a hot companion to the BSS. The temperature and radius of the BSS (T$_{\mathrm{eff}} = 8000^{+1000}_{-250}$ K, R/R$_\odot = 1.44 \pm 0.05$) derived from Gemini spectra and SED fitting using Kurucz atmospheric models are consistent with each other. The temperature and radius of the hotter companion of NH 84 (T$_{\mathrm{eff}} = 32,000 \pm 2000$ K, R/R$_\odot = 0.021 \pm 0.007$) derived by fitting Koester WD models to the SED suggest that it is likely to be a hot WD. The radial velocity derived from the spectra along with the proper motion from GAIA DR2 confirms NH 84 to be a kinematic member of the cluster. This is the second detection of a BSS-WD candidate in a GC, and the first in the outskirts of a low density GC. The location of this BSS in NGC 5466 along with its dynamical age supports the  mass-transfer pathway for BSS formation in low density environments. 
\end{abstract}


\section{Introduction}
Globular Clusters (GCs) are ideal laboratories to study the formation and properties of exotic interacting stellar systems such as Blue Stragglers (BSSs), X-ray Binaries, Cataclysmic Variables etc.  BSSs are stars located above the main sequence turn-off (MSTO) in the Color-Magnitude Diagram (CMD) \citep{Sandage}. They are brighter and bluer than the upper main sequence (MS) stars in the cluster (see e.g. \cite{Mirko2016}). The two main leading scenarios proposed for their formation are stellar collisions leading to mergers in high density environments \citep{Hills} and mass transfer (MT) from an evolved donor to a lower-mass star in a binary system in low density environments \citep{Mccrea, Chen}.\\ 
NGC 5466 is a metal poor ($[Fe/H]=-2.0$, \citep{Caretta2009} Galactic GC containing a large fraction of binaries ($\sim$6 $\%$) and BSS \citep{Beccari}. Being a low density GC (log$_{10}$ $\rho_{c} \sim$ 0.84 $L_{\odot}/pc^{3}$, \citep{Mclaugh2005}) as compared to other Galactic GCs of similar luminosity, mass-transfer is expected to be the dominant BSS formation mechanism where the primordial binaries can evolve in isolation in such environments \citep{Beccari}.

Ultra-Violet (UV) images are very effective in identifying BSS binaries with a hot companion as they show excess emission in the UV which, in general, is not expected from BSSs alone. Based on the FUV spectroscopy and Spectral Energy Distributions (SEDs) of 48 blue objects in 47 Tuc obtained with Hubble Space Telescope (HST), \cite{Knigge2008} discovered several interesting binary objects which also includes one BSS-WD binary in the cluster. \cite{Gosnell2014,Gosnell2015} detected white dwarf (WD) companions to seven BSSs in the open cluster NGC 188 based on Far-UV (F140LP, F150LP, and F165LP) observations with the HST. \cite{Subramaniam2016} detected  a hot companion (post-AGB/HB) to a BSS in NGC 188 using UVIT data on ASTROSAT, thus showing the importance of UV observations of BSSs.

Using a large sample of GC Color-Magnitude Diagrams (CMDs) from HST observations, \cite{Knigge2009} and \cite{Leigh2011} showed that the number of BSSs in the cores of GCs is strongly correlated with the total stellar mass of the core of the cluster. \cite{Leigh2007} and \citep{Leigh2013} found that there is little or no correlation of the BSS population with the collision rate in the core of the cluster, thus favouring binary evolution as the dominant channel for the formation of the BSSs. It was also found that the frequency of BSSs in GCs is correlated with the binary fraction \citep{Knigge2009, Milone2012}. 

\begin{table*}
\centering
\caption{Parameters of NGC 5466 used in this paper.}
\begin{tabular}{ccc}
\hline
Parameter & Value & Reference\\\hline
R.A. (J2000) & 14 05 27.29 & \cite{Goldsbury2010} \\
Dec (J2000) & +28 32 04.0 & \cite{Goldsbury2010} \\
$[Fe/H]$ & $-$2.0 dex & \cite{Caretta2009, Ferro}\\
Distance & 16.0 $\pm$ 0.6  kpc & \cite{Ferro}\\
Core radius, $r_{c}$ & $1\farcm43$ & \cite{Mclaugh2005}\\
Half-light radius, $r_{h}$ & $2\farcm3$ & \cite{Mclaugh2005}\\
$\mu_{RA}$ & $-5.404\pm0.004$ mas yr$^{-1}$ & \cite{Helmi2018} \\
$\mu_{Dec}$ & $-0.791\pm0.004$ mas yr$^{-1}$ & \cite{Helmi2018}\\\hline
\end{tabular}
\label{cluster}
\end{table*}

\begin{figure*}
\centering
\includegraphics[scale=0.37]{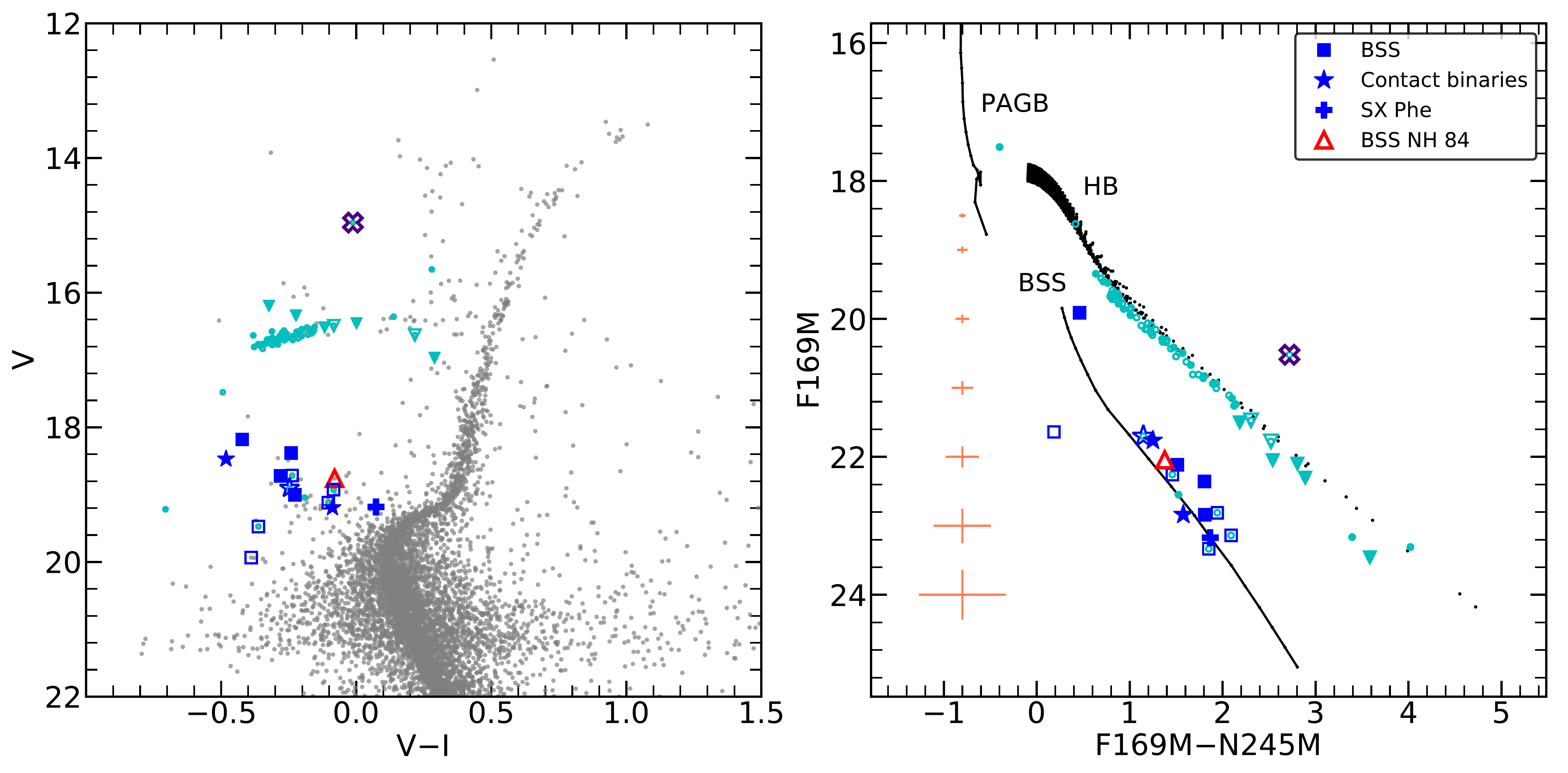}
\caption{F169M, (F169M$-$N245M) UV CMD (right panel) and its corresponding V, (V$-$I) optical CMD (left panel) with BSSs detected in all the UVIT filters. Open symbols are UVIT cross-matched ground (CFHT) detections and closed symbols are the UVIT cross-matched HST detections. Cyan dots are the objects detected down to F169M = 24 mag. The cross symbol is an anomalous Cepheid whereas the lower triangles are RR lyrae variables. The UV CMD is over plotted with a Padova model isochrone (black line and dots) of age 12.6 Gyr and metallicity [Fe/H]= $-$1.98 \citep{Caretta2009}.}
\label{cmd}
\end{figure*}

\cite{Ferraro2009} discovered two BSS sequences in the optical CMD of GC M30, suggesting that the redder ones arise from the evolution of close binaries that are still experiencing MT which was in agreement with binary evolution models. Another explanation for the two BSS sequences in M30 was given by \cite{Jiang2017} where they showed that binary evolution contributes to the formation of BSS in both sequences. Thus, identification of BSSs with hot companions using UV observations are crucial to understanding their formation mechanism in binary systems. 

In this paper, we present the SED analysis of a BSS candidate of GC NGC 5466 based on UVIT and Gemini observations. We outline the observations and data reduction in Section \ref{obs}, the UV CMD in Section \ref{uvcmd}, spectroscopic analysis in Section \ref{specdata}, SED of the BSS in Section \ref{bssed}, followed by discussion, summary and conclusions in Sections \ref{discus} and \ref{sum}.\\

\section{Observations and Data Reduction}
\label{obs}

\begin{figure}
\centering
\includegraphics[width=\columnwidth]{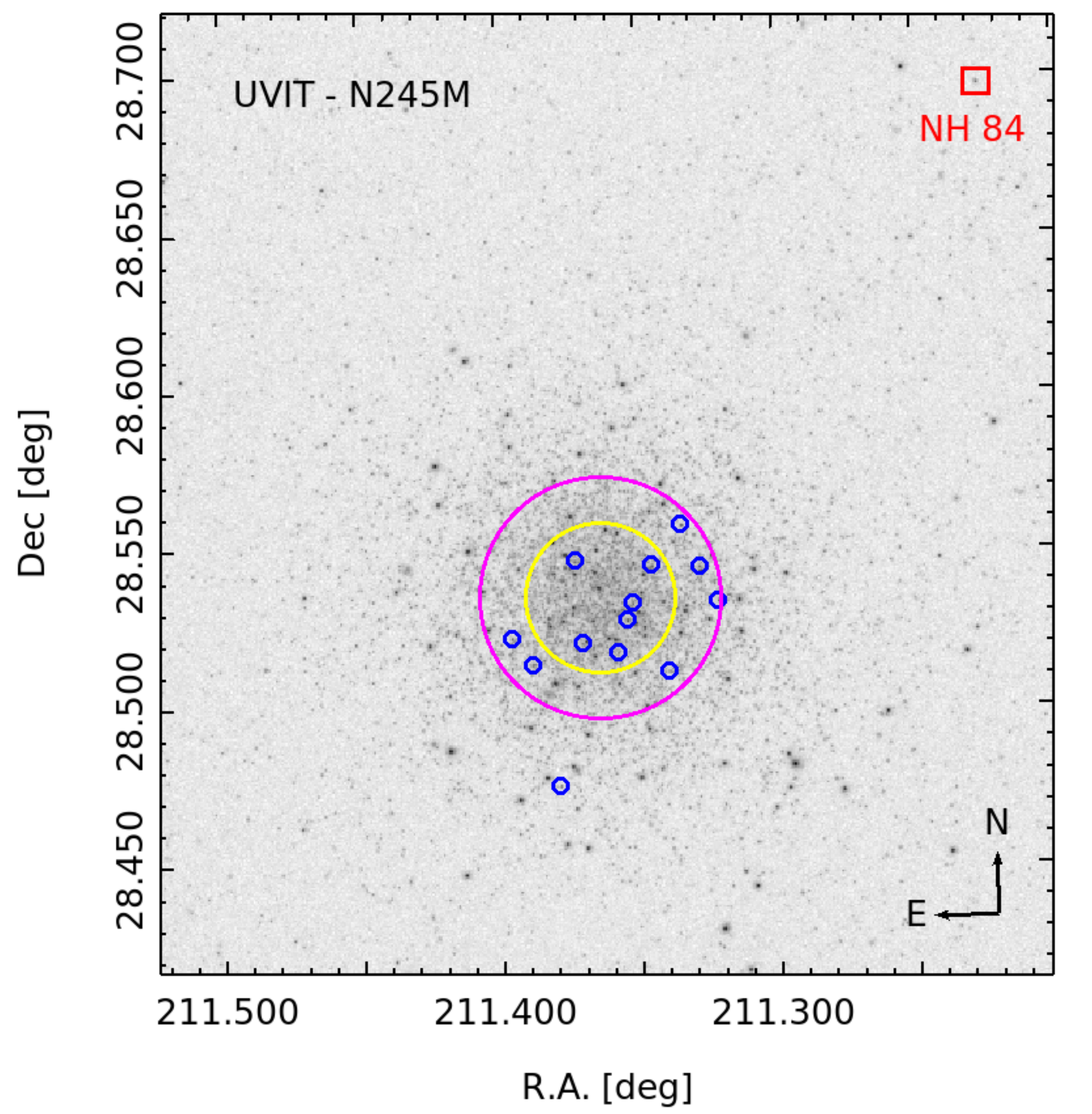}
\caption{14 BSSs detected by UVIT are overlaid on top of an N245M filter image of UVIT. The red square symbol in the figure is the BSS target in our study. The yellow and magenta circles are the core radius ($\sim 1\farcm43$) and half-light radius ($\sim 2\farcm3$) of the cluster from the center \citep{Mclaugh2005}.}
\label{image}
\end{figure}

The cluster was observed with the UVIT telescope as a part of the GT proposal (G05$\_$009) during 3-4 June, 2016. UVIT is one of five payloads onboard AstroSat, which is operated by the Indian Space Research Organization (ISRO). The calibration of the instrument can be found in \cite{Tandon2017a}. Full details of the telescope and instrument is available in \cite{Tandon2017b} and \cite{Subramaniam_c2016}.

The data were acquired in four filters of UVIT- two in the FUV (F148W and F169M) and two in NUV channels (N245M and N263M). The images were processed using the CCDLAB software \citep{Postma2017} which corrects for the satellite drift, flat field and distortion. Isolated stellar sources in the UVIT images have FWHM $\sim$1\farcs5 and $\sim$1\farcs2 in the FUV and NUV channels, respectively. In terms of angular resolution, the UVIT images are thus far superior to those from GALEX (4\farcs5-5\farcs5). 
Crowded-field photometry was performed using DAOPHOT/IRAF tasks and packages \citep{Stetson}. Aperture and saturation corrections were done to obtain the final magnitudes in the AB system, details can be found in \cite{Tandon2017a}. The photometric errors in magnitude for all the UVIT filters are found to be within 0.4 down to 24th magnitude.\\

\section{UV Color-Magnitude Diagram}
\label{uvcmd}
We cross-matched UVIT data with HST-ACS survey data of the GC NGC 5466 \citep{Sarajedini2007} for the central regions with FOV 3$\farcm4 \times 3\farcm4$ and ground based data provided by Peter Stetson for the region beyond the FOV of HST. We separated them into various stellar populations such as HB and BSS based on their locations in both the optical and UV CMDs. The parameters of the cluster adopted in this study are given in Table \ref{cluster}.

\begin{figure}
\centering
\includegraphics[width=\columnwidth]{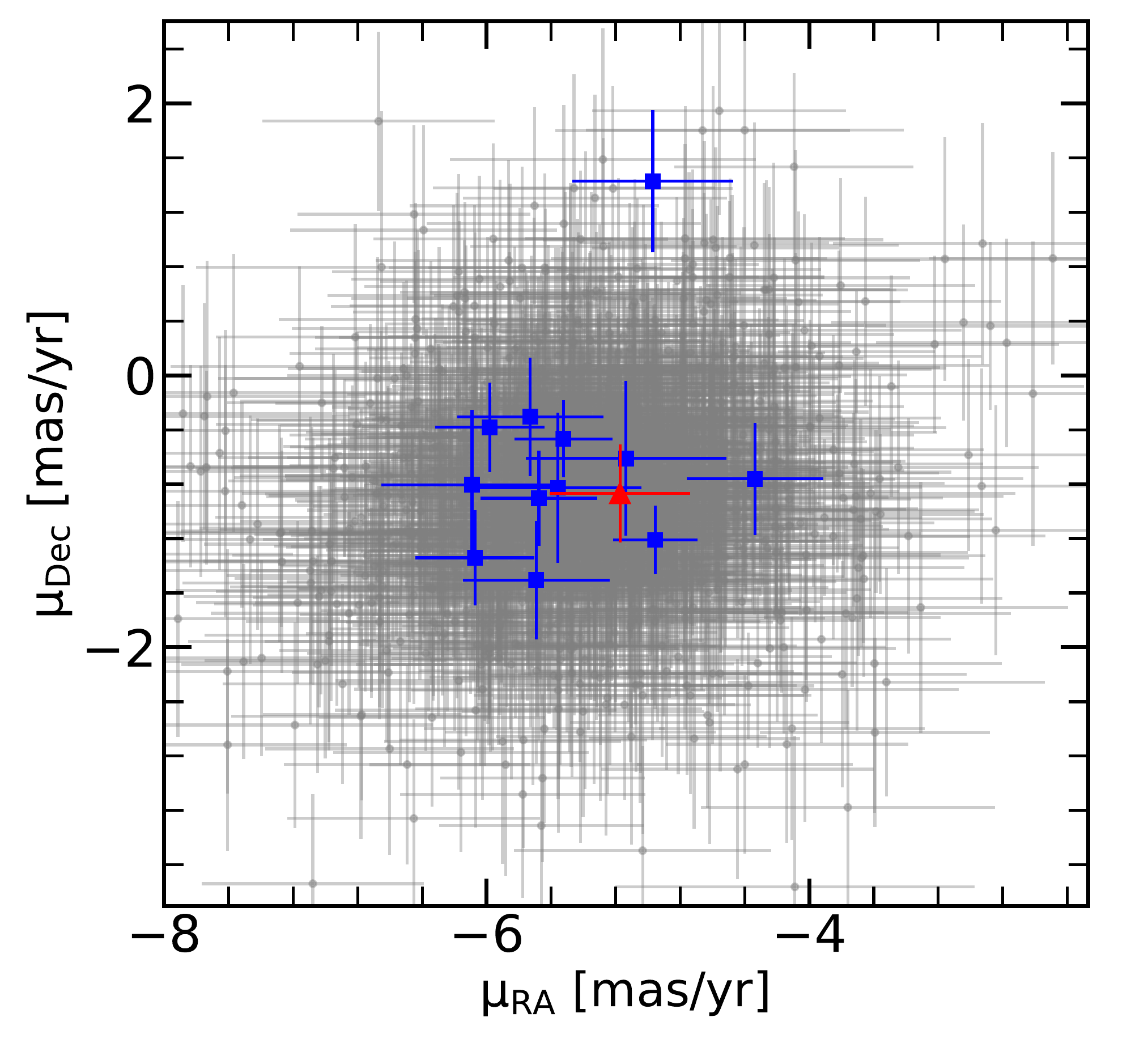}
\caption{Vector-point diagram of the 14 BSSs (blue squares) detected by UVIT relative to other cluster members (grey) in the catalog given by \cite{Helmi2018}. The red triangle in the figure is BSS NH 84.}
\label{bspm}
\end{figure}

To check their cluster membership, we used the GAIA DR2 Proper Motion (PM) catalog of the cluster NGC 5466 given by \cite{Helmi2018}. Their catalog consists of the list of PM member stars of the cluster where the procedure for the member selection is described in detail in Appendix A.1 of their paper. The vector-point diagram of BSSs (blue squares) relative to other cluster members (grey dots) in the PM catalog is shown in Figure \ref{bspm} where we clearly notice that the FUV detected BSSs are grouped around the mean PM derived by \cite{Helmi2018} (Table \ref{cluster}) except BSS NH 48. According to the HST PM study by \cite{Mirko2016}, this BSS is a PM member. In total, we found 14 BSSs detected in all the UVIT filters to be PM members. In addition, 63 HB stars are also found to be PM members. The typical uncertainties in the PMs are $\sim$ 0.12 and 0.42 mas yr$^{-1}$ for the HB and BSSs respectively. These 14 BSSs are marked in the FUV (F169M) and NUV (N245M) images as shown in Figure \ref{uvit_bss}, where we can clearly see that the BSS are spatially resolved and PSF photometry can be successfully performed.

\begin{figure*}
\centering
\includegraphics[scale=0.255]{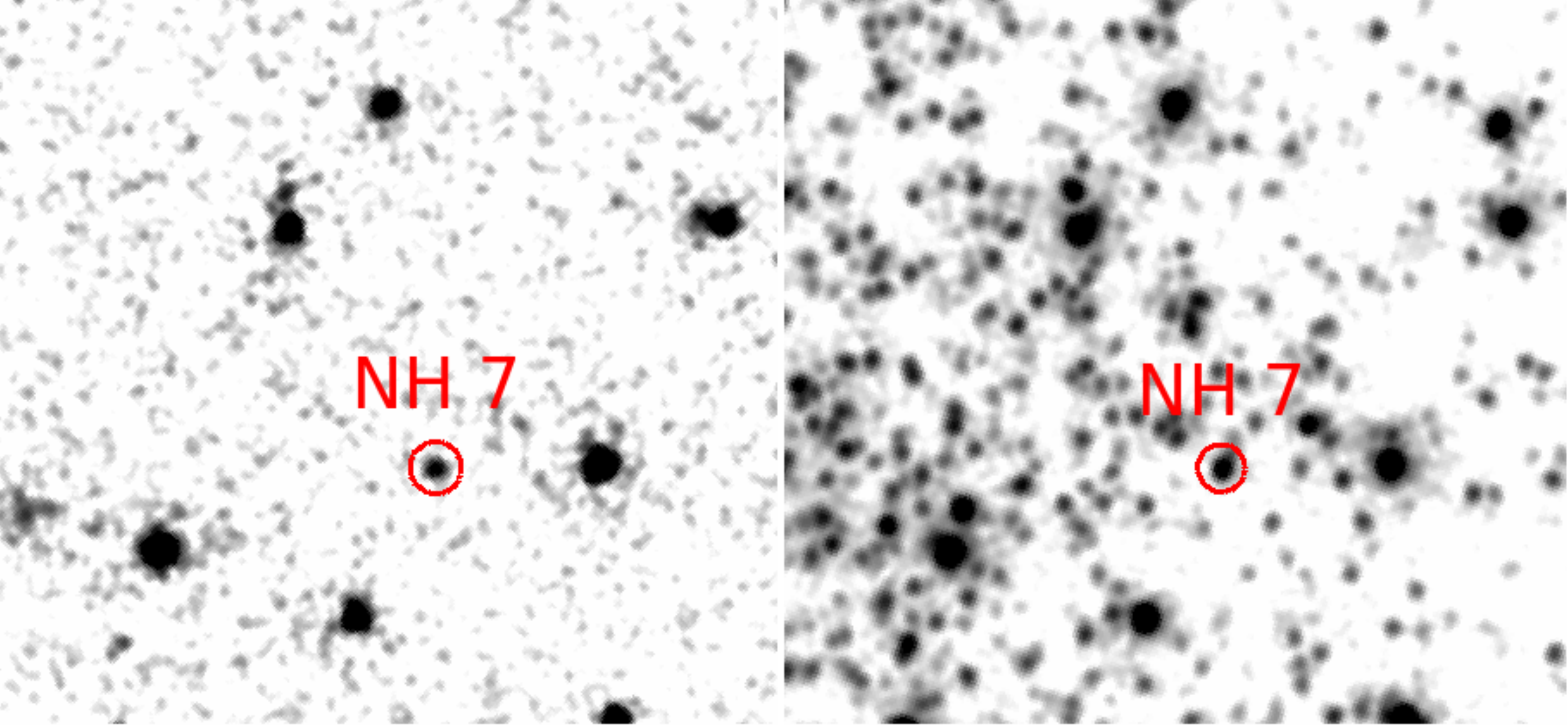}
\hspace{0.1cm}
\vspace{0.2cm}
\includegraphics[scale=0.255]{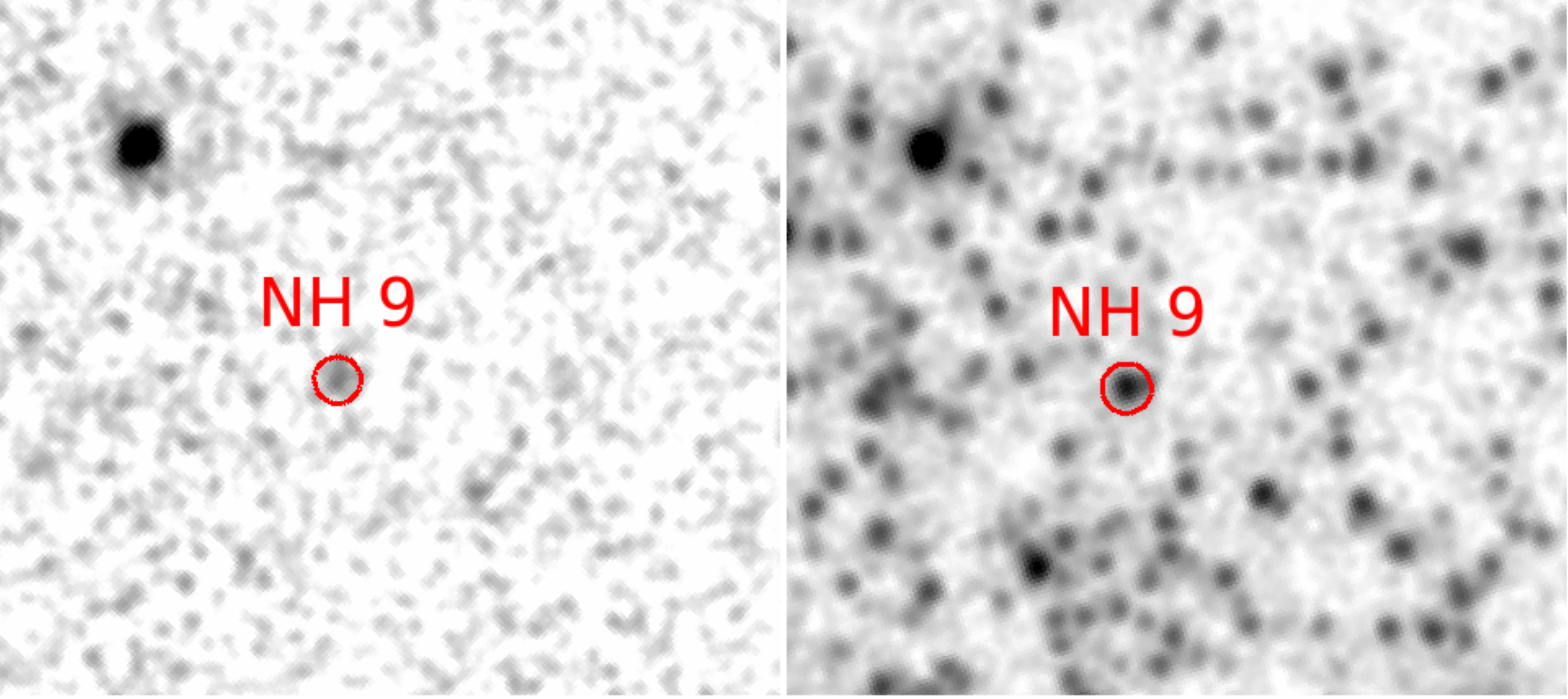}\\
\includegraphics[scale=0.25]{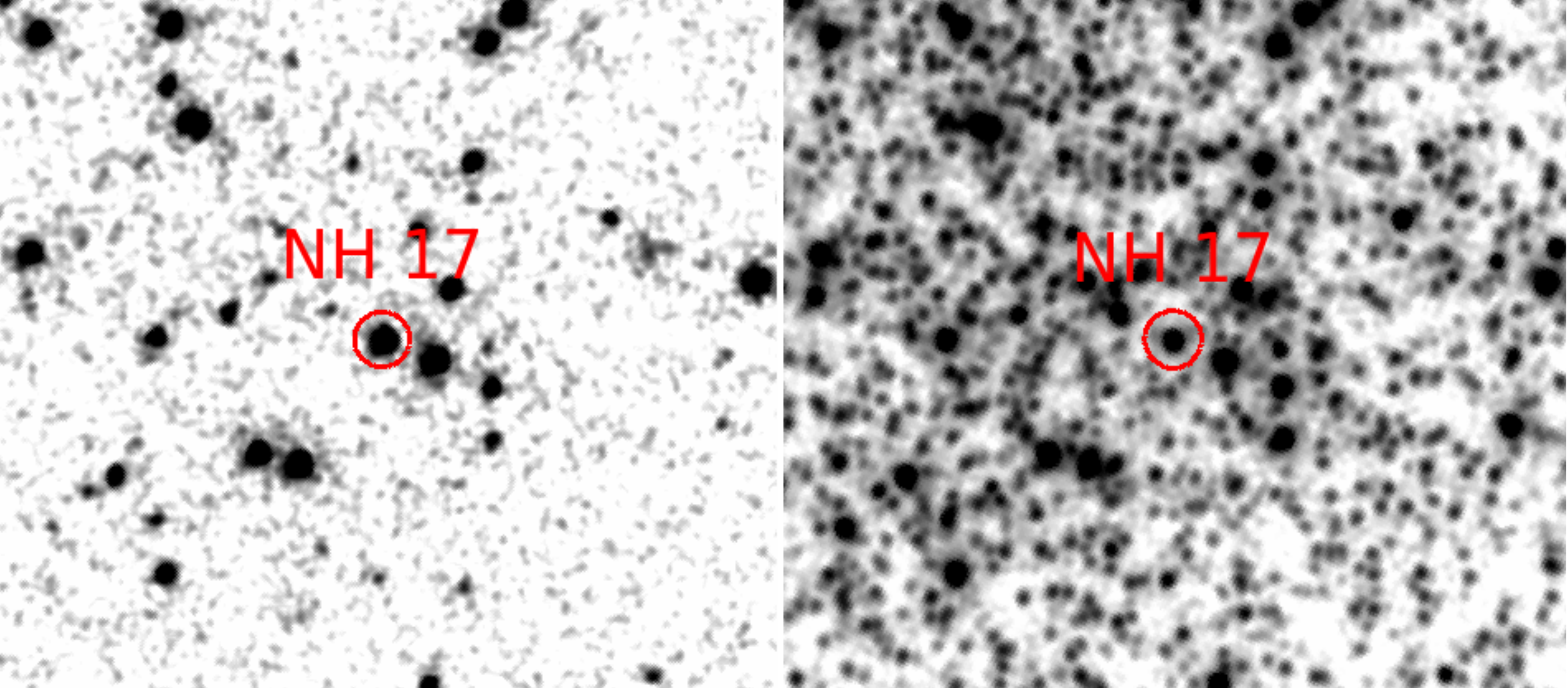}
\hspace{0.1cm}
\vspace{0.2cm}
\includegraphics[scale=0.25]{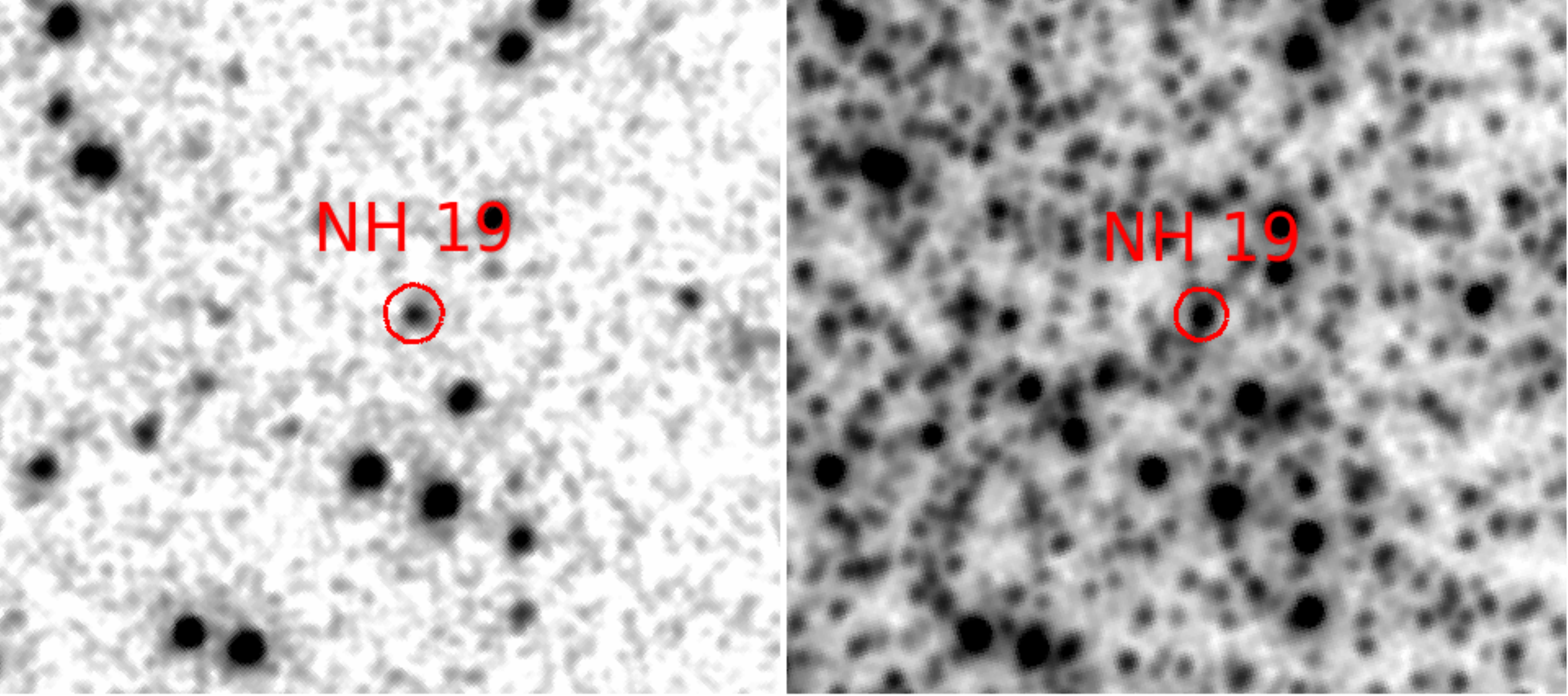}\\
\includegraphics[scale=0.25]{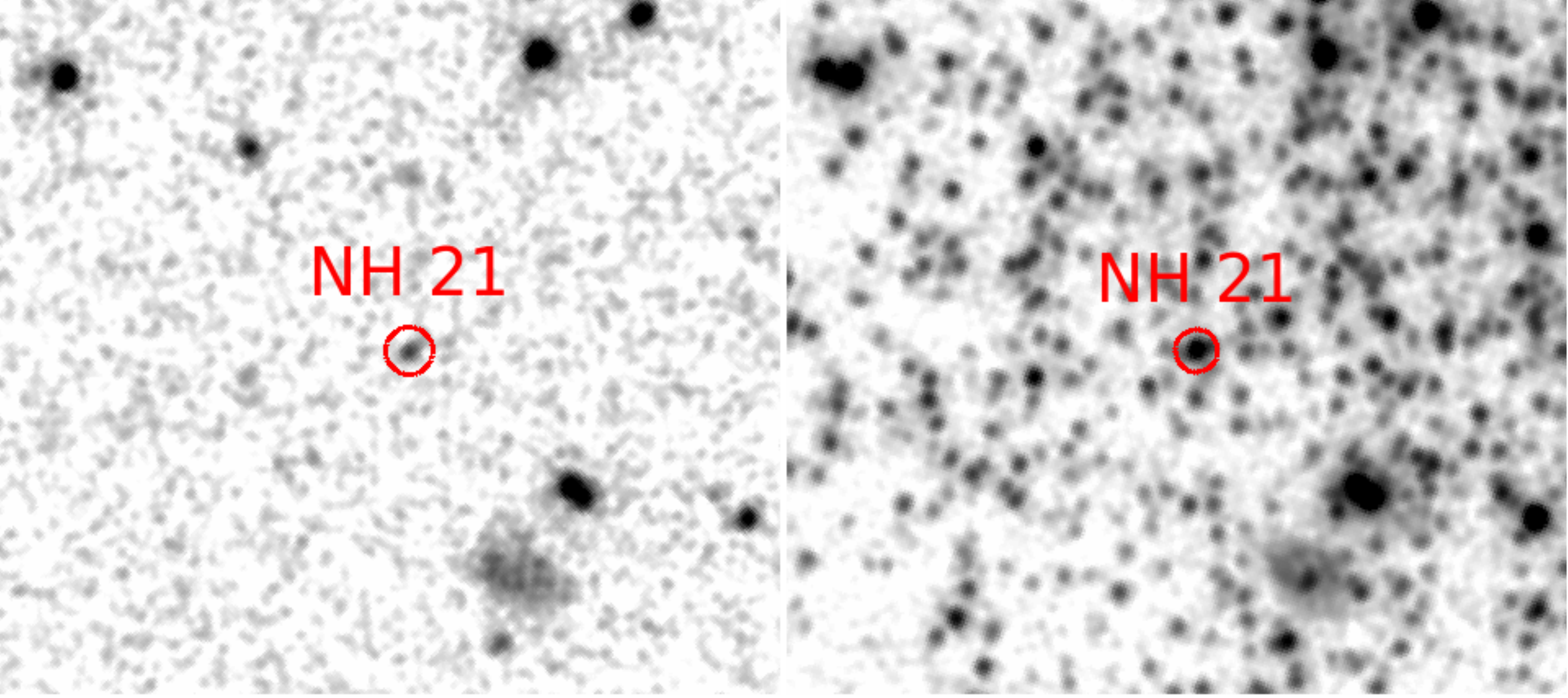}
\hspace{0.1cm}
\vspace{0.2cm}
\includegraphics[scale=0.25]{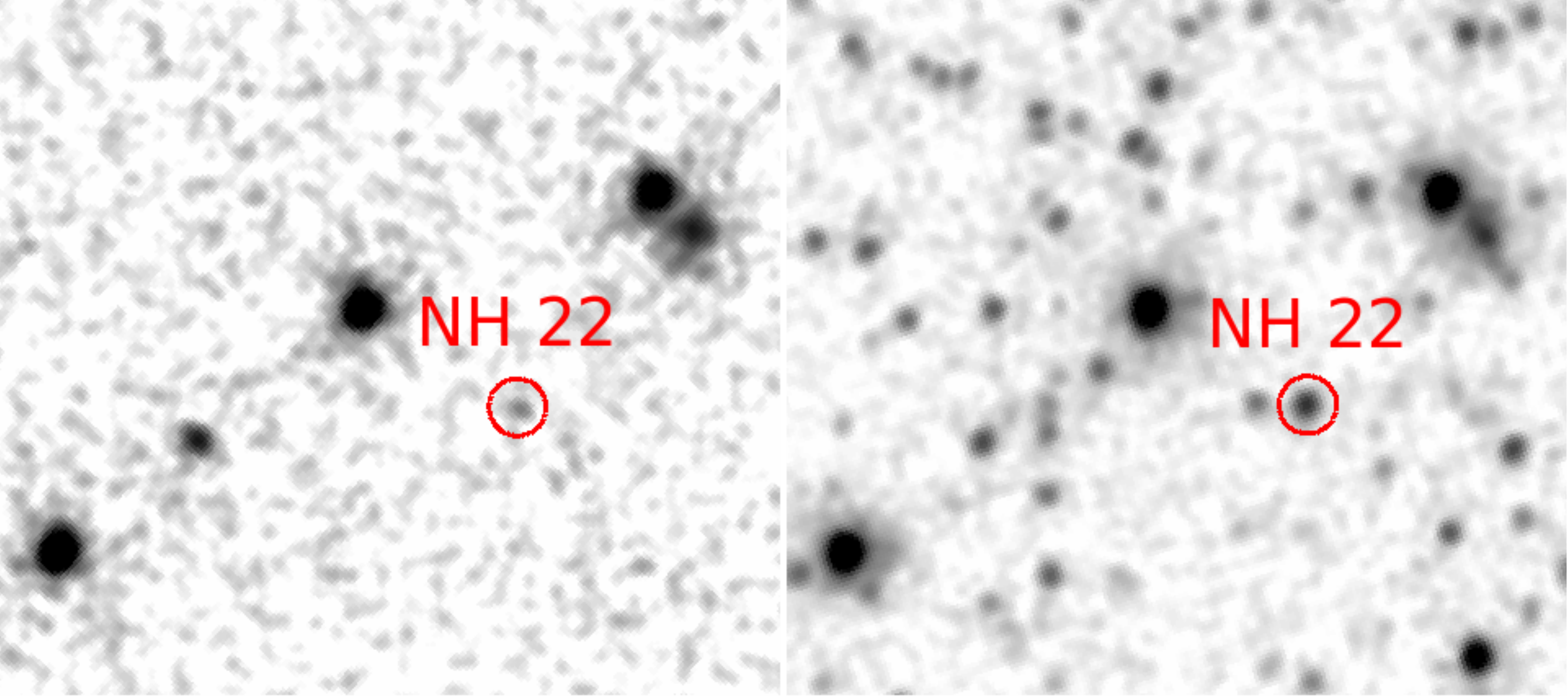}\\
\includegraphics[scale=0.25]{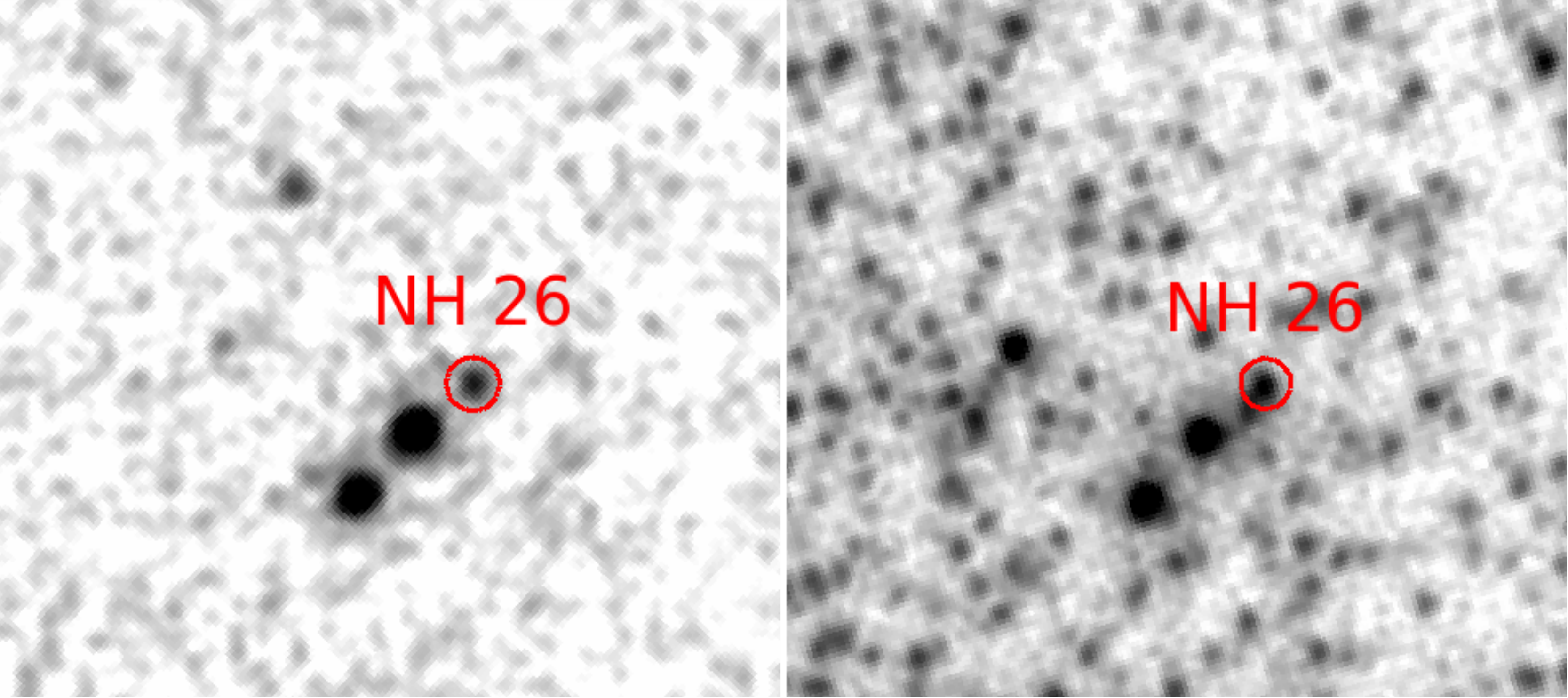}
\hspace{0.1cm}
\includegraphics[scale=0.25]{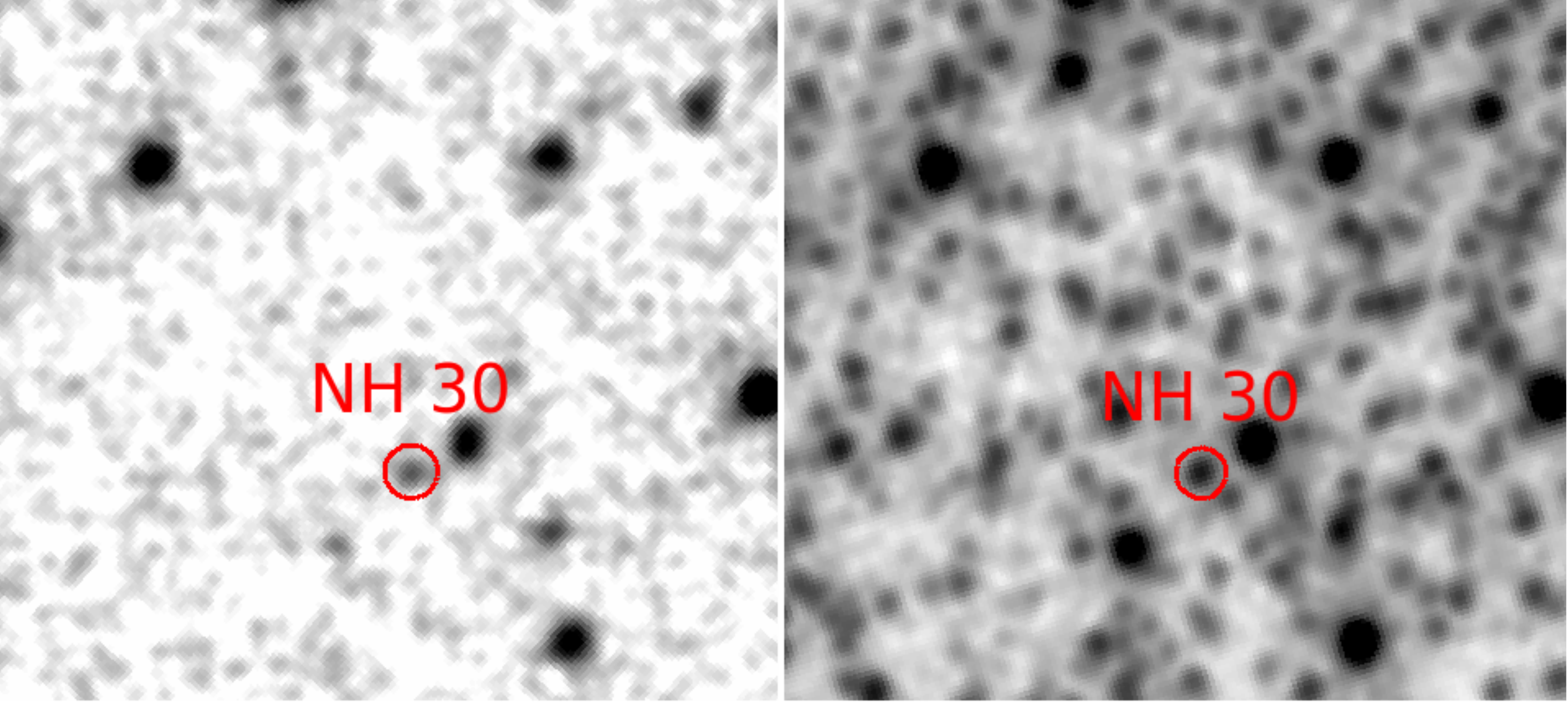}\\
\caption{Location of BSSs on FUV-F169M (Left) and NUV-N245M (Right) images of UVIT.}
\label{uvit_bss}
\end{figure*}

\begin{figure*}
\centering
\includegraphics[scale=0.25]{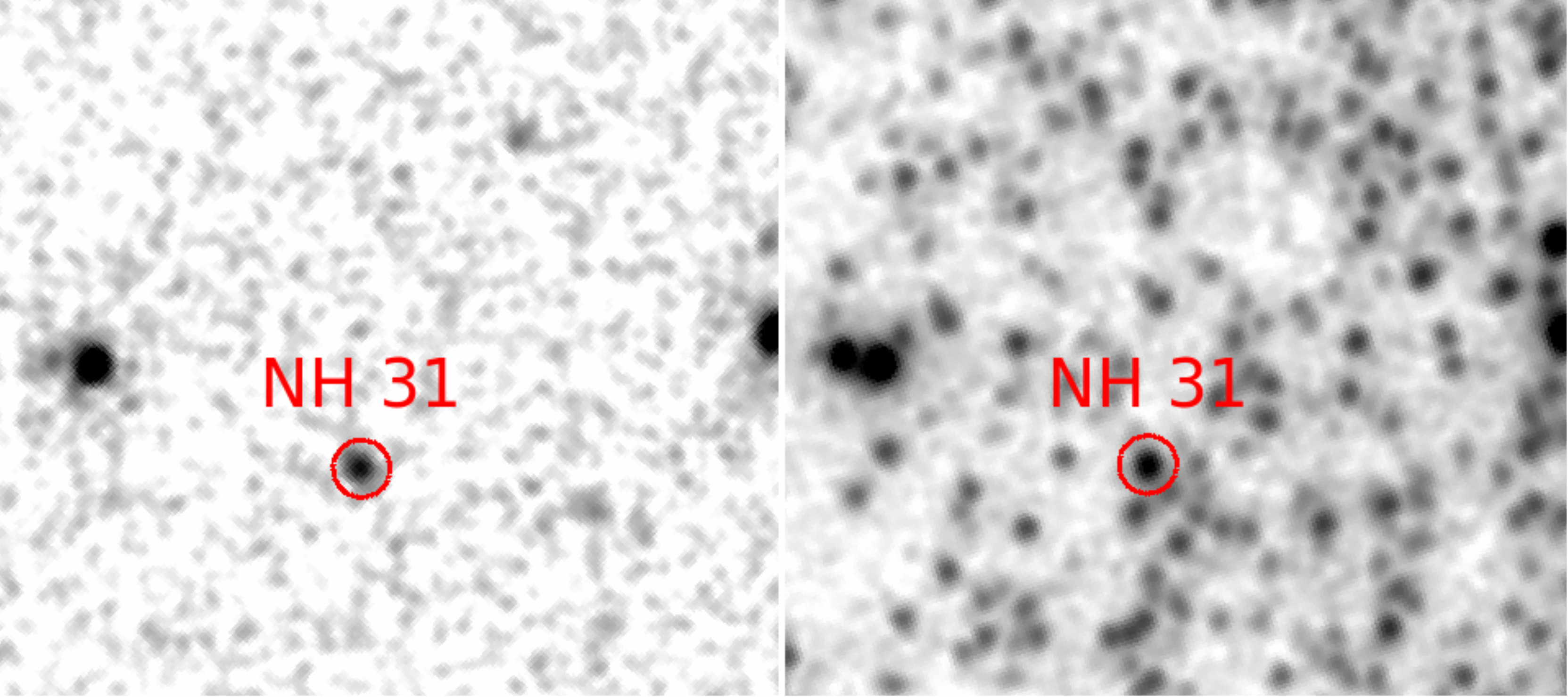}
\hspace{0.1cm}
\vspace{0.2cm}
\includegraphics[scale=0.25]{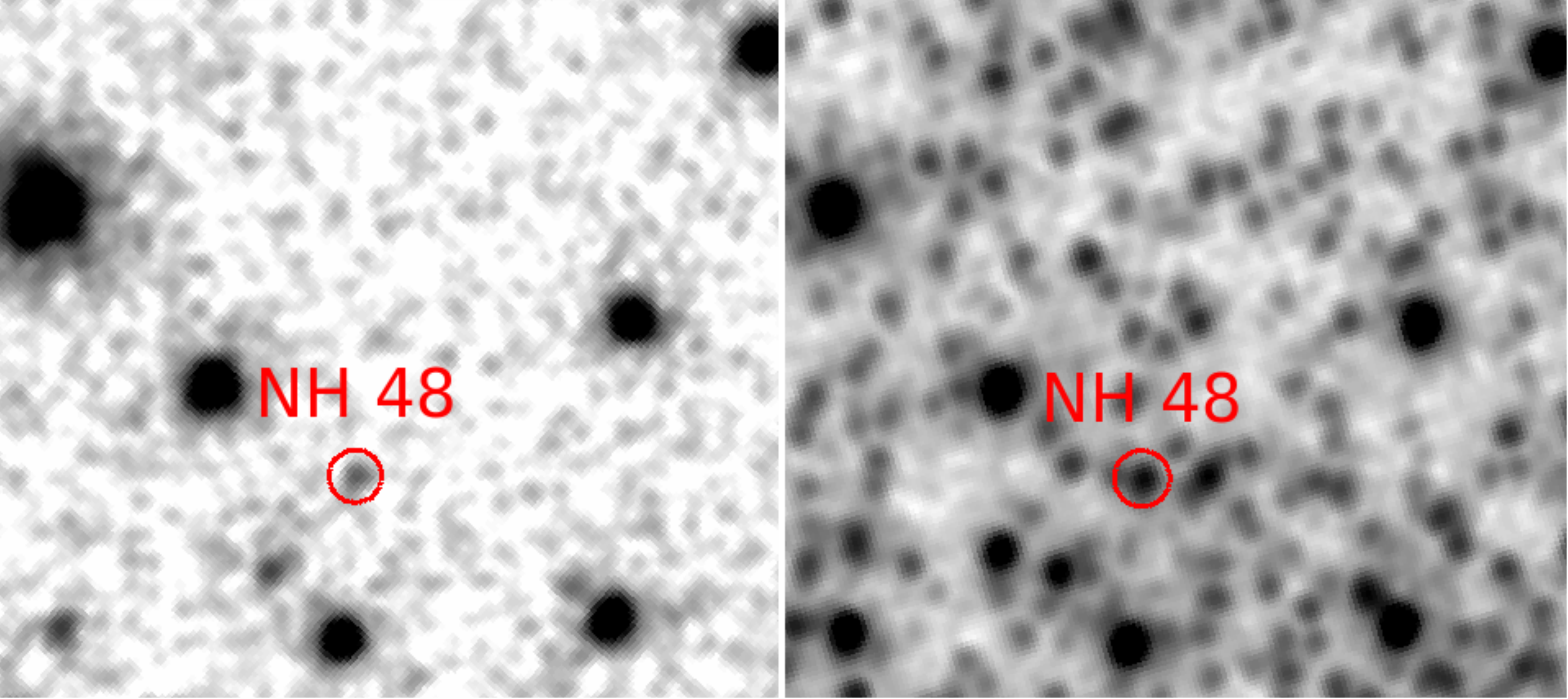}\\
\includegraphics[scale=0.25]{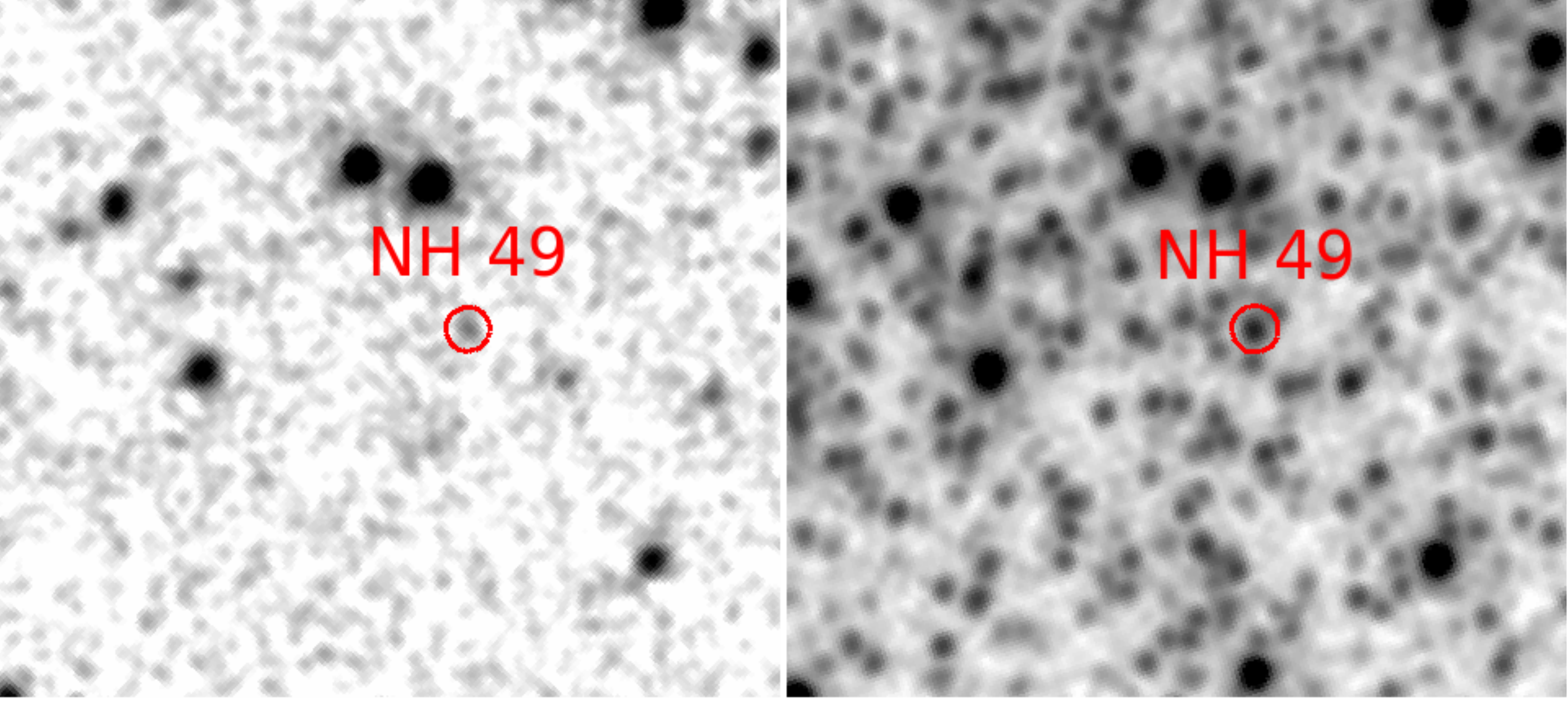}
\hspace{0.1cm}
\vspace{0.2cm}
\includegraphics[scale=0.25]{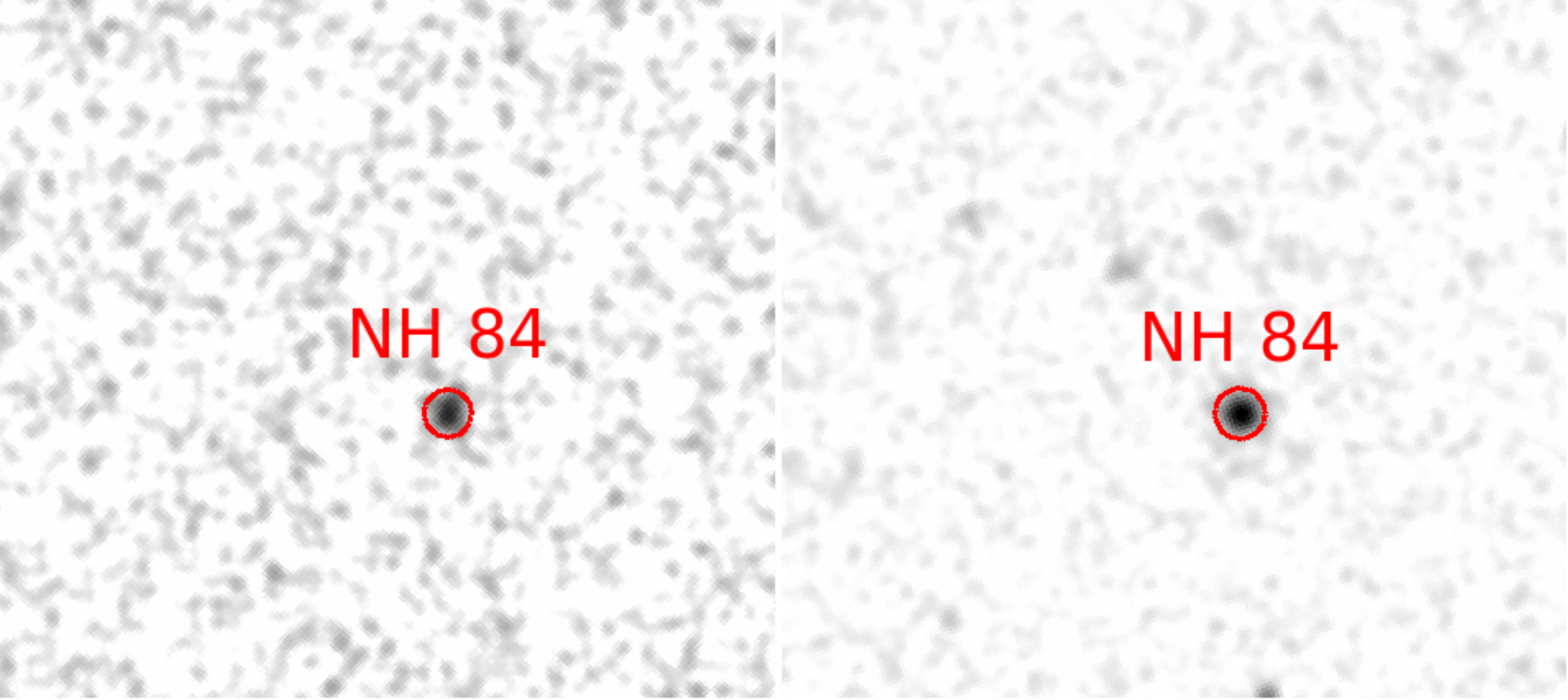}\\
\includegraphics[scale=0.25]{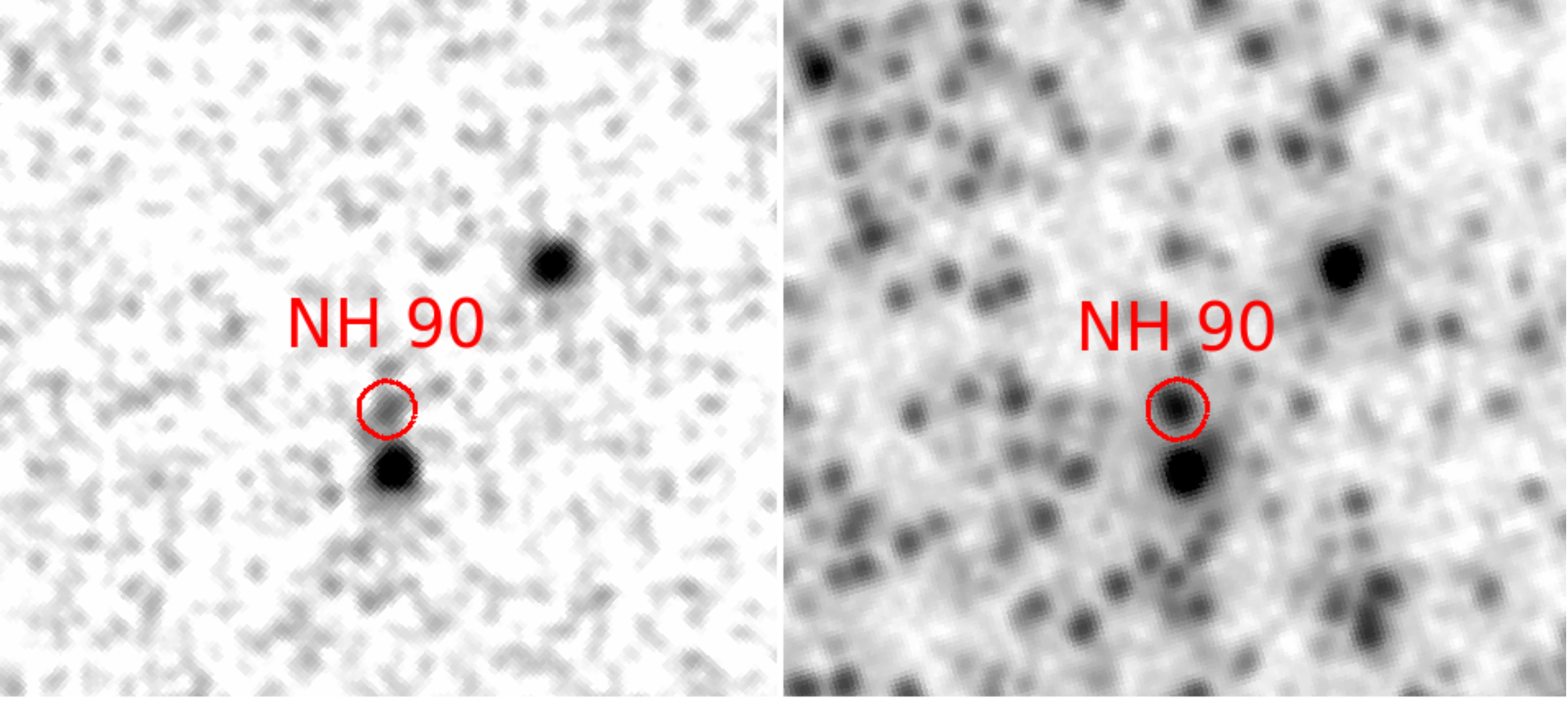}
\hspace{0.1cm}
\includegraphics[scale=0.25]{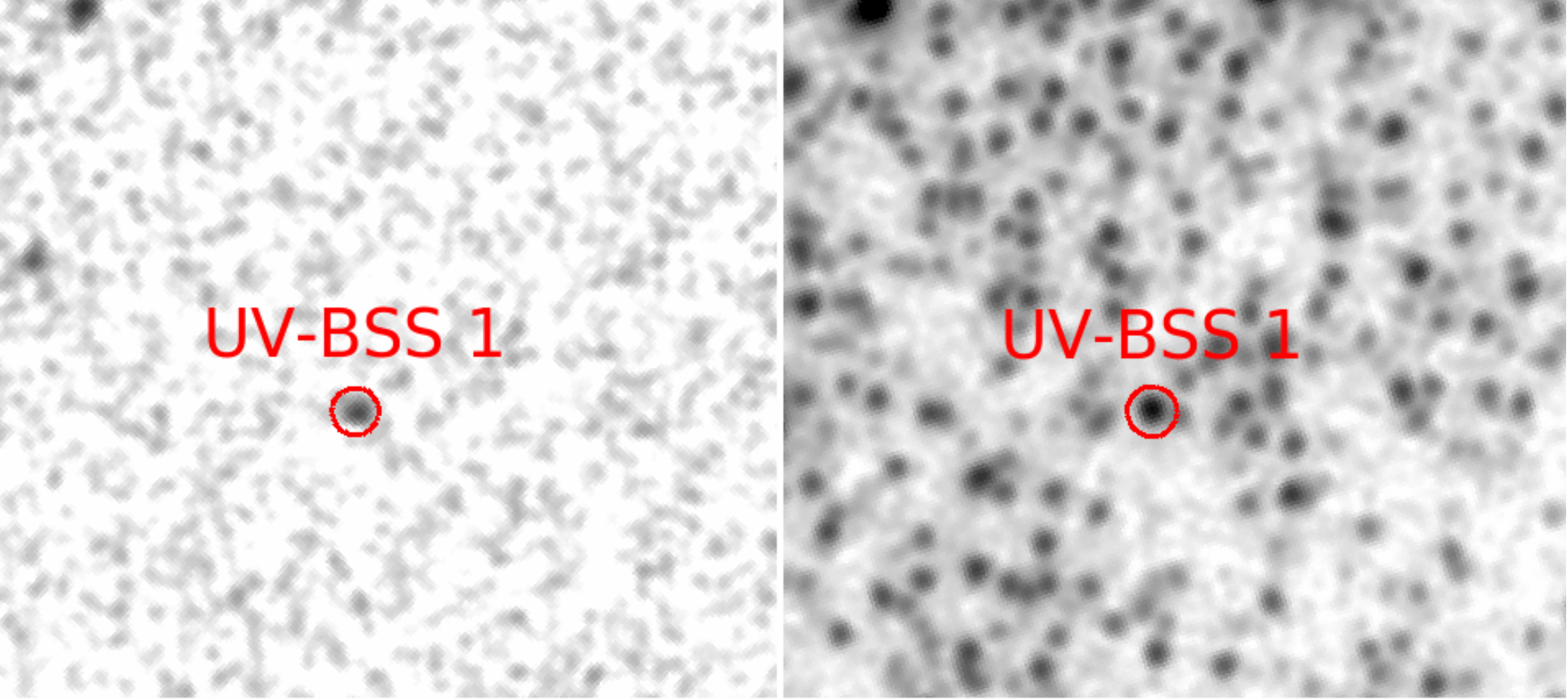}\\
\caption{Continuation of Figure \ref{uvit_bss}}
\end{figure*}

The PM cleaned F169M vs (F169M$-$N245M) CMD (right panel) along with the optical CMD (left panel) are shown in Figure \ref{cmd} where the detected BSSs are shown as blue squares. The UV CMD is over plotted with a Padova model \citep{Marigo2007, Marigo2008} isochrone (black line and dots) of age 12.6 Gyr, metallicity [Fe/H]= $-$1.98 \citep{Caretta2009} for a distance modulus of 16.0 \citep{Ferro}. The isochrone is generated by convolving Padova models with UVIT filter response curves \citep{Tandon2017a} using the Flexible Stellar Population Synthesis (FSPS) code \citep{Conroy, Conroy2010}. The FSPS generates a locus of BSS, assuming them to be MS stars with masses in excess of the turn-off mass, which uniformly populates 0.5 magnitudes above the MSTO to 2.5 magnitudes brighter than the MSTO as shown in the figure. Reddening is not corrected as it is close to zero, E(B$-$V) = 0.00 \citep{Zinn1985}.
The cross symbol in the figures is a known Anomalous Cepheid \citep{Zinn1976}. The bright star in the UV CMD at F169M, F169M$-$N245M $\sim$ 17.5, -0.4, lying close to the PAGB evolutionary model track is classified as a AGB-manqu\'e star by \cite{Schiavon2012}.

\cite{Sandquist} presented BVI photometry of Red Giants and BSSs based on their observations with the KPNO 0.9m telescope. They identified 94 BSS candidates based on their locations in optical CMDs. Of the 14 BSSs detected by UVIT which are listed in Table \ref{bss}, 13 BSSs were found to be in common with their catalog \citep{Sandquist} suggesting that we have an additional BSS named as UV-BSS1 (Table \ref{bss}). The GALEX magnitudes of the BSSs given in the table are obtained by performing PSF photometry on the GALEX FUV and NUV intensity maps (OBJECT ID- GI1\_056017\_NGC5466). We used the BSS nomenclature from \cite{Nemec1987} for our study.

Among the UV detected BSSs, NH 17 is the brightest BSS as shown in the left panel of the Figure \ref{cmd}. NH 19, NH 30 are known W-UMa type contact binaries and NH 31 is an eclipsing binary \citep{Mateo}. BSS NH 49 is a known SX Phe variable \citep{Jeon2004}. One of the BSS from the \cite{Sandquist} catalog, NH 87, is found to be bluer than the BSS model track in the UV CMD (empty blue box in right panel of Figure \ref{cmd}) whereas other BSSs are distributed around the track. This BSS does not have the PM information from GAIA DR2. We obtained the GMOS-N spectra of this object (see Section~\ref{specdata}).

The location of the FUV detected BSSs from Table \ref{bss} are shown as blue circles in Figure \ref{image} overlaid on UVIT's N245M filter image of the cluster. The red square in the figure is BSS NH 84. We found that most of the BSSs (12 out of 14) are located inside the half-light radius of the cluster.

\begin{table*}
\centering
\caption{UV magnitudes of the FUV detected BSS candidates in all the UVIT and GALEX filters that are PM members \citep{Helmi2018}. The UV-BSS1 does not have a counterpart in the BSS catalog of \cite{Sandquist}.}
\setlength{\tabcolsep}{7pt}
\scalebox{0.63}{%
\begin{tabular}{cccccccccccc}
\hline
& &	&\multicolumn{5}{c}{UVIT} & \multicolumn{2}{c}{GALEX} & \multicolumn{2}{c}{Proper motion}\\
BSS ID	&	RA (J2000) 	&	DEC (J2000) & r	&	F148W	&	F169M	&	N245M	&	N263M & FUV & NUV & $\mu_{RA}$ & $\mu_{Dec}$\\
&	[deg]	&	[deg]	& [$\arcmin$]	& [mag]	&	[mag]	&	[mag]	&	[mag]	& [mag] & [mag] & [mas yr$^{-1}$] & [mas yr$^{-1}$] \\\hline
UV-BSS	1	&	211.3393	&	28.51122	&	1.83	&	23.08	$\pm$ 	0.30	&	22.55	$\pm$ 	0.19	&	21.02	$\pm$ 	0.06	&	21.09	$\pm$ 	0.09	&	23.24	$\pm$ 	0.20	&	21.13	$\pm$ 	0.07	&	-5.63	$\pm$ 	0.58	&	-0.69	$\pm$ 	0.49	\\
NH	7	&	211.32196	&	28.53361	&	2.22	&	22.88	$\pm$ 	0.30	&	22.26	$\pm$ 	0.18	&	20.80	$\pm$ 	0.05	&	20.63	$\pm$ 	0.08	&	22.63	$\pm$ 	0.15	&	20.81	$\pm$ 	0.04	&	-5.98	$\pm$ 	0.34	&	-0.38	$\pm$ 	0.33	\\
NH	9	&	211.33525	&	28.55766	&	2.06	&	23.15	$\pm$ 	0.31	&	23.14	$\pm$ 	0.28	&	21.05	$\pm$ 	0.07	&	21.65	$\pm$ 	0.18	&	-	 	 	&	21.10	$\pm$ 	0.04	&	-6.09	$\pm$ 	0.56	&	-0.80	$\pm$ 	0.55	\\
NH	17	&	211.35456	&	28.52768	&	0.64	&	20.05	$\pm$ 	0.08	&	19.91	$\pm$ 	0.06	&	19.45	$\pm$ 	0.03	&	19.43	$\pm$ 	0.04	&	19.96	$\pm$ 	0.08	&	19.56	$\pm$ 	0.04	&	-4.95	$\pm$ 	0.26	&	-1.21	$\pm$ 	0.25	\\
NH	19\textsuperscript{a}	&	211.35267	&	28.53305	&	0.6	&	21.66	$\pm$ 	0.16	&	21.76	$\pm$ 	0.12	&	20.51	$\pm$ 	0.05	&	19.97	$\pm$ 	0.10	&	21.68	$\pm$ 	0.11	&	20.30	$\pm$ 	0.06	&	-6.07	$\pm$ 	0.37	&	-1.34	$\pm$ 	0.35	\\
NH	21	&	211.38859	&	28.51381	&	1.79	&	22.39	$\pm$ 	0.20	&	22.36	$\pm$ 	0.16	&	20.55	$\pm$ 	0.05	&	20.37	$\pm$ 	0.09	&	22.60	$\pm$ 	0.18	&	20.33	$\pm$ 	0.04	&	-5.52	$\pm$ 	0.31	&	-0.47	$\pm$ 	0.28	\\
NH	22	&	211.37913	&	28.4753	&	3.64	&	23.59	$\pm$ 	0.35	&	23.33	$\pm$ 	0.26	&	21.48	$\pm$ 	0.09	&	21.47	$\pm$ 	0.13	&	-	        	 	&	21.59	$\pm$ 	0.08	&	-5.14	$\pm$ 	0.62	&	-0.61	$\pm$ 	0.57	\\
NH	26	&	211.34568	&	28.54491	&	1.15	&	22.23	$\pm$ 	0.18	&	22.12	$\pm$ 	0.17	&	20.60	$\pm$ 	0.07	&	20.39	$\pm$ 	0.08	&	-	 	 	&	20.61	$\pm$ 	0.09	&	-5.69	$\pm$ 	0.46	&	-1.51	$\pm$ 	0.44	\\
NH	30\textsuperscript{a}	&	211.37085	&	28.52038	&	0.92	&	23.32	$\pm$ 	0.36	&	22.84	$\pm$ 	0.20	&	21.26	$\pm$ 	0.09	&	21.66	$\pm$ 	0.14	&	-	 	 	&	-	 	 	&	-5.73	$\pm$ 	0.45	&	-0.30	$\pm$ 	0.44	\\
NH	31\textsuperscript{a}	&	211.39603	&	28.52215	&	1.84	&	22.15	$\pm$ 	0.22	&	21.70	$\pm$ 	0.13	&	20.55	$\pm$ 	0.06	&	20.48	$\pm$ 	0.06	&	21.82	$\pm$ 	0.12	&	20.42	$\pm$ 	0.07	&	-5.68	$\pm$ 	0.36	&	-0.90	$\pm$ 	0.35	\\
NH	48	&	211.37318	&	28.54648	&	0.87	&	23.17	$\pm$ 	0.33	&	22.84	$\pm$ 	0.25	&	21.03	$\pm$ 	0.07	&	20.97	$\pm$ 	0.13	&	-	 	 	&	21.06	$\pm$ 	0.07	&	-4.97	$\pm$ 	0.50	&	1.43	$\pm$ 	0.52	\\
NH	49\textsuperscript{b}	&	211.35802	&	28.51746	&	1.07	&	23.68	$\pm$ 	0.39	&	23.17	$\pm$ 	0.25	&	21.30	$\pm$ 	0.07	&	21.31	$\pm$ 	0.10	&	23.53	$\pm$ 	0.34	&	-	 	 	&	-5.56	$\pm$ 	0.52	&	-0.83	$\pm$ 	0.55	\\
NH	90	&	211.3281	&	28.54438	&	1.92	&	22.84	$\pm$ 	0.28	&	22.81	$\pm$ 	0.25	&	20.87	$\pm$ 	0.09	&	20.56	$\pm$ 	0.10	&	-	 	 	&	-	 	 	&	-5.10	$\pm$ 	0.38	&	-1.32	$\pm$ 	0.35	\\
NH	84	&	211.22649	&	28.69703	&	12.15	&	22.47	$\pm$ 	0.25	&	22.04	$\pm$ 	0.14	&	20.67	$\pm$ 	0.06	&	20.39	$\pm$ 	0.09	&	22.28	$\pm$ 	0.13	&	20.68	$\pm$ 	0.03	&	-5.17	$\pm$ 	0.44	&	-0.87	$\pm$ 	0.36	\\\hline
%\hline
\label{bss}
\end{tabular}
}
\\\textsuperscript{a}Eclipsing and contact binaries \citep{Mateo}
\\\textsuperscript{b}SX Phe variable \citep{Jeon2004, Ferro}\\
\end{table*}

\section{GMOS-N spectroscopic data}
\label{specdata}
We obtained spectroscopic data using the GMOS-N spectrograph mounted on the 8.1-meter Gemini-North telescope for two sources detected in FUV filters, namely NH 87 and NH 84 (see Table \ref{bss}). The observations were part of the Gemini program GN-2018A-FT-113 (PI: M. Simunovic) and were taken during June, 2018. We used the R400\_G5305 grating and a 0\farcs75 long slit which yielded a dispersion of 0.074 nm/pix and a spectral resolution R $\approx$ 1300 for the $\sim$ 460-900 nm spectral range.~We took 4$\times$330 sec exposures in each central wavelength (700 and 705 nm) in order to cover the GMOS-N detector chip gaps. The data was reduced using standard IRAF routines available in the Gemini/GMOS package which resulted in flux-calibrated spectra at a signal-to-noise of $\sim$60, shown in Figure \ref{specsed} for NH 84.

The spectra of NH 87 showed a flat continuum and strong emission lines consistent with an HII region, suggesting it to be a star-forming galaxy at redshift z $\sim$ 0.09. This also supports the previous classification by SDSS of this object as a galaxy. Hence we discard this object as a contaminant and focus on the spectroscopic analysis of NH 84, which is confirmed as a stellar source. 

\subsection{Radial velocity and spectral fitting of NH 84}
\label{spec_fit}
\begin{figure}
\centering
\includegraphics[width=\columnwidth]{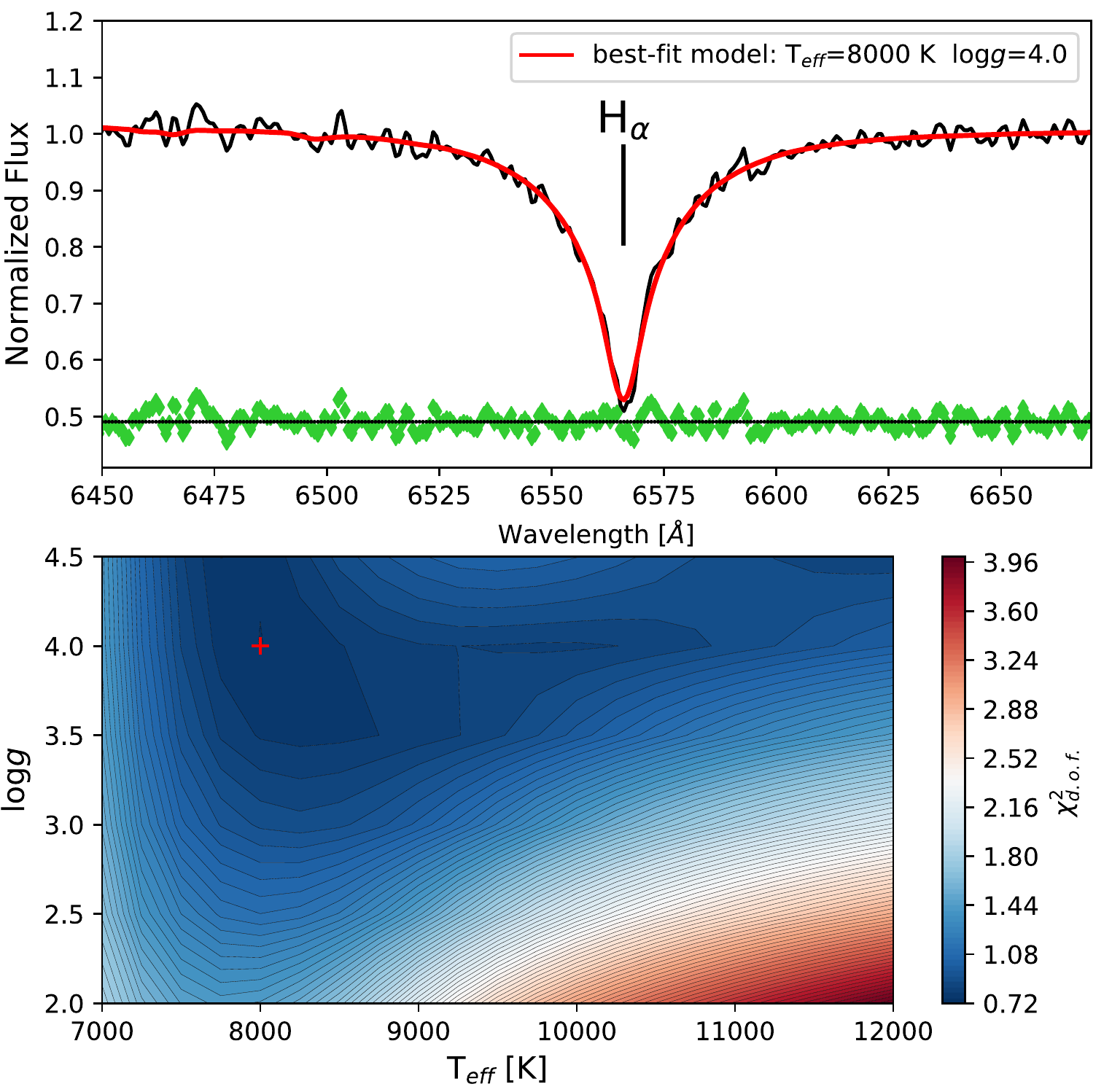}
\caption{Top: GMOS-N spectrum of NH 84 around the H$_{\alpha}$ line. The best-fit model is shown in the red solid line and the residual values in green points. Bottom: Color map and contour plot of the reduced $\chi^2$ as a function of T$_{\mathrm{eff}}$ and log $g$. The best-fit value is marked as a red cross in the upper left corner.}
\label{Halpha_fit}
\end{figure}

The stellar parameters T$_{\mathrm{eff}}$ and log $g$ were obtained by fitting the shape of the H$_\alpha$ line, which is commonly used as a T$_{\mathrm{eff}}$ and log $g$ indicator that is also independent of rotational broadening. The spectral fitting method was a $\chi^2$ minimization using the pPXF python package \citep{capellari17} with a grid of synthetic spectra from the Coehlo library \citep{coelho14}, fixed at [Fe/H] = $-2.0$ dex. The grid was limited to T$_{\mathrm{eff}}$ values between 7000$-$12000\,K in 250\,K steps, whereas log $g$ were taken between 2.0$-$5.0 in 0.5 steps. The synthetic spectra were then degraded to the spectral resolution of the GMOS-N data and the pPXF spectral fitting was performed allowing only a radial velocity shift and no kinematic broadening. The observed spectrum of NH 84 and best-fit model are shown in Figure~\ref{Halpha_fit}. We obtained T$_{\mathrm{eff}}$ = 8000 K and log $g$ = 4.0 for the best-fit parameters of NH 84.~As it can be seen in the lower panel of Figure~\ref{Halpha_fit}, the distribution of $\chi^2$ is not uniformly distributed around the minimum, and hence asymmetric uncertainties are present.~To obtain robust uncertainties, we take the best-fit synthetic model and add random gaussian noise such that its signal-to-noise=60, as in the observed data, and run pPXF to obtain the best-fit model of this artificial data sample. We run 1000 iterations and obtain probability distributions for the best-fit parameters. This way, we adopt T$_{\mathrm{eff}}$ = 8000$^{+1000}_{-250}$ K and log $g$ = 4.0$^{+0.5}_{-0.5}$ as the uncertainties, obtained from the parameter distribution interval that contains 95\% of the probability, as found with our Monte Carlo approach.

We used the Fourier cross-correlation method to derive the radial velocity of NH 84. The data were cross-correlated against the best-fit synthetic spectra using the FXCOR routine in IRAF. The measured heliocentric radial velocity for NH 84 is $v_{helio} = 128 \pm 30$ km s$^{-1}$, which is consistent with previous measurements of the systemic radial velocity of NGC\,5466 found in the literature. \cite{Harris} reports 110.7 km s$^{-1}$, while \cite{shetrone2010} measured a weighted average value of 118.0 $\pm$ 0.4 km s$^{-1}$ for 67 stars, and \cite{lamb2015} obtains an average value of 121.05 km s$^{-1}$ from 3 stars in NGC\,5466. Hence our results are consistent with NH 84 being a kinematic cluster member.

\section{SED of BSS NH 84}\label{bssed}
In order to understand the multi-wavelength energy budget of the FUV detected BSS, we generated their SEDs and estimated the temperature, luminosity and radius. We used the virtual observatory tool, VOSA (VO SED Analyzer, \cite{Bayo}) for SED analysis. VOSA calculates synthetic photometry for a selected theoretical model using filter transmission curves. It performs a $\chi^{2}$ minimization test by comparing the synthetic photometry with observed data to get the best fit parameters of the SED. We estimated the reduced $\chi^{2}$ value using the expression given by
\begin{equation}
\small
    \chi^{2}_{red} =\frac{1}{N-N_{f}} \sum_{i=1}^{N}\Big\{\frac{(F_{o,i}-M_{d}F_{m,i})^{2}}{\sigma_{o,i}^{2}}\Big\}
\label{chi2}
\end{equation}
where N is the number of photometric data points, N$_{f}$ is the number of free parameters in the model, $F_{o,i}$ is the observed flux, $M_{d}F_{m,i}$ is the model flux of the star, $\displaystyle{M_{d}=\bigg(\frac{R}{D}\bigg)^{2}}$ is the scaling factor corresponding to the star (where R is the radius of the star, and D is the distance to the star) and $\sigma_{o,i}$ is the error in the observed flux. We used Kurucz stellar atmospheric models \citep{Castelli, Castelli2004} for the BSSs which covers the UV to IR wavelength range. The model's free parameters are [Fe/H], T$_{eff}$ and log $g$. We fixed the metallicity [Fe/H] = $-$2.0, close to the cluster metallicity and varied the other two parameters ($T_{eff}$ and log $g$) in the Kurucz models to fit the SED of the BSSs.

\begin{figure}
\includegraphics[width=\columnwidth]{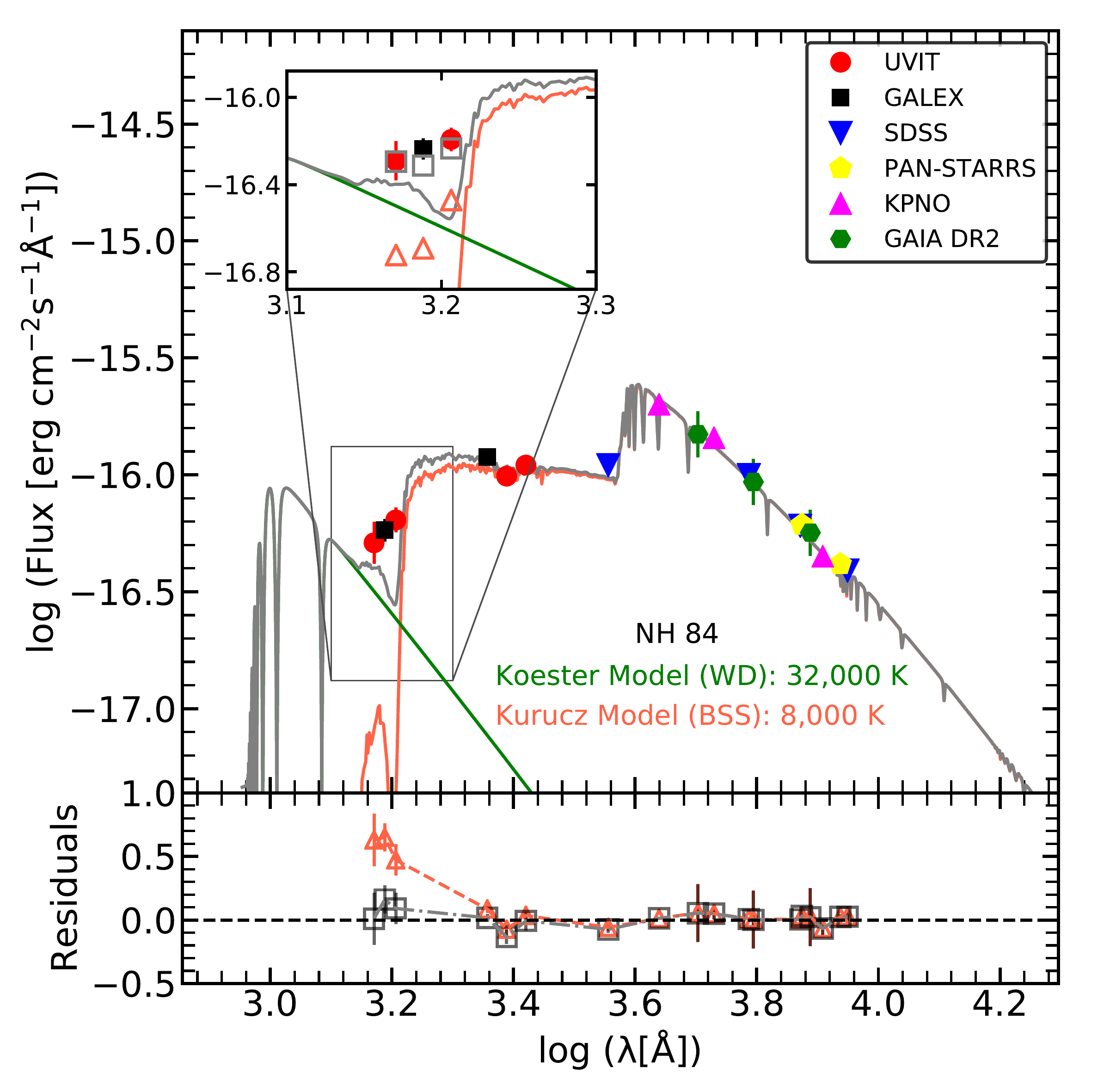}
\caption{SED of NH 84 with a composite spectra (gray color) consisting of Kurucz model (light-red color) and Koester WD model (green color). The zoomed in plot shows the FUV part of the SED fitted with a single and composite spectrum where the light-red empty triangles indicate the Kurucz synthetic flux and grey empty squares indicate the combined synthetic flux. The residuals obtained with a single and composite spectrum fit are shown as light-red empty triangles and grey empty squares in the lower panel. See Section \ref{bssed} for details.}
\label{sednh84}
\end{figure}

The SED of the BSS NH 84 was constructed by combining the flux measurements of UVIT (4 passbands) with GALEX (FUV and NUV), GAIA DR2 (3 passbands) \citep{Gaia2016,Gaia2018}, KPNO (3 passbands), SDSS (4 passbands) \citep{sdss2012}, and PAN-STARRS (2 passbands) \citep{pan3-2016} surveys (upper panel, Figure \ref{sednh84}) obtained from VO photometry. The number of photometric points used for constructing the SED of NH 84 is 16. The UV flux measurements of NH 84 along with the exposure times are given in Table \ref{uvflux}. We found that fitting the full SED with a single Kurucz model spectrum of T$_{eff}$ = 8000 $\pm$ 125 K and log $g$ = 4.0 $\pm$ 0.5 resulted in a large $\chi_{red}^{2} \sim $ 5.75 for the given degrees of freedom (14). This is clear from the residual plot shown in the lower panel of Figure \ref{sednh84}, which shows the difference between the observed flux and the synthetic flux normalised with respect to the observed flux, corresponding to the flux measurements in each passband. We find that the residual plot shows a rise in flux in the UV wavelengths for a single spectrum fit (shown as light-red empty triangles in the figure). Similarly, we checked the SEDs and the residual plot of other 13 BSSs. If we find the residual to be more than 50\% in the FUV wavelengths, we classify the BSS as having UV excess. We found that out of 14 BSSs, there are 6 BSSs which show UV excess. Our focus is on BSS NH 84 in this study as we have the radial velocity membership confirmation from our spectroscopic study (Section \ref{specdata}).

\begin{table}
\centering
\caption{UV Flux measurements of BSS NH 84}
\setlength{\tabcolsep}{0.5pt}
%\resizebox{.5\textwidth}{!}{
\begin{tabular}{cccc}
\hline
Filter & Exposure time & Flux & Flux Error \\
& sec & erg cm$^{-2}$ s$^{-1}$  $\mathrm{\AA^{-1}}$ & erg cm$^{-2}$ s$^{-1}$  $\mathrm{\AA^{-1}}$\\\hline
\multicolumn{4}{c}{UVIT}\\\hline
F148W & 2609 & 5.12E-17 & 1.06E-17 \\
F169M & 6036 & 6.41E-17 & 7.96E-18 \\
N245M & 6091 & 9.87E-17 & 5.65E-18 \\
N263M & 4258 & 1.10E-16 & 8.29E-18 \\\hline
\multicolumn{4}{c}{GALEX}\\\hline
FUV & 1838 & 5.80E-17 & 6.45E-18 \\	
NUV & 3529 & 1.19E-16 & 3.79E-18 \\\hline
\label{uvflux}
\end{tabular}
\end{table}

\begin{figure}
\includegraphics[width=\columnwidth]{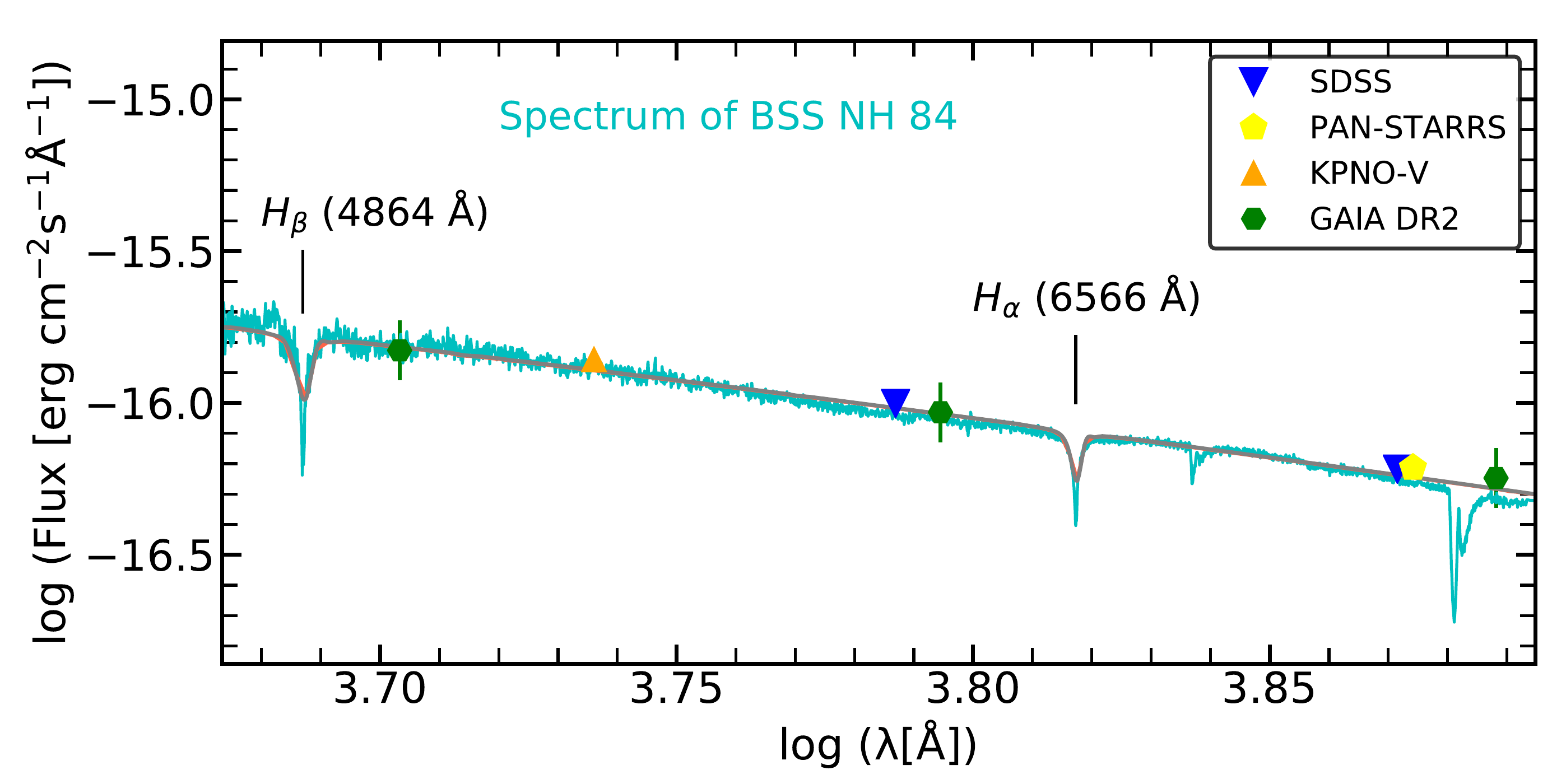}
\caption{Gemini spectrum of NH 84 (cyan color) overplotted on the SED of NH 84. The wavelengths corresponding to H$_{\alpha}$ and H$_{\beta}$ absorption lines of the BSS are marked in the figure. Note that the absorption features redward of the H$_{\alpha}$ line are telluric bands.}
\label{specsed}
\end{figure}

In order to address the UV-excess found in BSS NH 84, we generated a composite spectrum by combining the fluxes of Kurucz models for BSS \citep{Castelli, Castelli2004} and Koester WD models \citep{Tremblay2009} for the hot component. We independently obtained the SED fit parameters of the BSS  using Kurucz models for a fixed metallicity ([Fe/H] = $-$2.0) considering wavelengths longer than 2000 $\AA$ and found that it is in agreement with the parameters obtained from spectra (Section \ref{spec_fit}). The SED fit parameters for the BSS are given in Table \ref{sedpar}. The Gemini spectrum is consistent with the photometric flux measurements and the best fit composite model (grey line), as shown in Figure \ref{specsed}. Note that the observed absorption features redward of the H$_{\alpha}$ line are telluric bands. Note also that the shown model (grey line) comes from very low resolution spectral models, which explains why the Balmer line shapes are not well matched, as compared to Figure \ref{Halpha_fit}. We found T$_{eff}$ = 8000 K as the best fit value for the cool component from both the SED and the spectrum. Keeping the BSS parameters fixed, we varied the parameters of the Koester WD models assuming a log $g$ = 7.5, to get the best fit combination for the full SED as given in Table \ref{sedpar}. 

The UV excess part of the SED fitted with a single Kurucz spectrum and composite spectrum are shown in a zoomed in plot of the SED (upper panel of Figure \ref{sednh84}) where the light-red empty triangles indicate the single-component Kurucz synthetic flux and grey empty squares indicate the combined synthetic flux (Kurucz + Koester) in the respective FUV filters. When inspecting the zoomed in panel in Figure \ref{sednh84}, the reader should focus their attention on the synthetic flux points (light-red empty triangles and grey empty squares) when comparing to the observed data points, instead of comparing to the solid-line model spectra, which give the misleading impression of a bad fit. Large residuals found for single spectrum fit reduce to almost zero with the composite spectrum fit, in particular for the residuals in the UV filters (shown as grey empty squares in the lower panel of Figure \ref{sednh84}). Thus, NH 84 is found to have a hotter WD companion of temperature 32,000 $\pm$ 2000 K. The $\chi^{2}_{red}$ value for the composite spectrum of NH 84 is $\sim$ 1.62 corresponding to a 95\% confidence level. We note that the largest non-zero residuals are still at the far blue end, where the WD fit is supposed to compensate. 

We estimated the basic parameters (luminosities and radii) of the components of BSS NH 84 using the values (T$_{eff}$, M$_{d}$) obtained from the SED fitting. For estimating the radii of the components, we used the relation of M$_{d}$ as mentioned in Equation \ref{chi2} by adopting a distance of 16 $\pm$ 0.6 kpc \citep{Ferro}. The radius of the cool component of BSS NH 84 is $\sim$ 1.44 $\pm$ 0.05 R$_{\odot}$ whereas that of the hot component is $\sim$ 0.021 $\pm$ 0.001 R$_{\odot}$ which is close to the typical radii of WDs \citep{Tremblay}. The uncertainties in the radii are estimated using the equation $\displaystyle{\Delta R=\frac {R\Delta D} {D}}$ where, $\Delta D$ = 0.6 kpc taken from \cite{Ferro}. We calculated the luminosities of the components of the BSS using the relation:

\begin{equation}
    \frac{L}{L_{\odot}}= \bigg(\frac{R}{R_{\odot}}\bigg)^{2}  \bigg(\frac{T_{\mathrm{eff}}}{T_{\odot}}\bigg)^{4}
\end{equation} 
which are given in Table \ref{sedpar}. The hot component has a luminosity of $\sim$ 0.42 $\pm$ 0.11 L$_{\odot}$ whereas the cool component has $\sim$ 7.58 $\pm$ 1.10 L$_{\odot}$.  

\subsection{Uncertainties in the WD parameters}
\label{error}
In order to evaluate the upper limit of uncertainty found in T$_{eff}$ estimate of the BSS from spectroscopy (Section \ref{specdata}), we checked the SED fits for the BSS temperatures ranging from 8000 - 9000 K using Kurucz models. We found that fitting the SED of the BSS with T$_{eff}$ = 8250 K has less $\chi^{2}_{red}$ than that of 8000 K. Though, it brings down the residuals in the UV wavelengths to 30$\%$ from 50$\%$, but the individual $\chi^{2}$ for the passband near the Balmer jump (KPNO B) increases with increasing T$_{eff}$. As the Balmer jump is very sensitive to T$_{eff}$ of the cooler component, 8000 K is  more appropriate for the T$_{eff}$ of the BSS from the SED fits. We also note that, if we consider a T$_{eff}$ of 8250 K for the BSS, then the best fitting parameters using Koester models (T$_{eff}$ and R) are 30,000 K and 0.014 R$_{\odot}$ which are also consistent with the hot component being a WD. The total $\chi^{2}_{red}$ value increases for temperatures larger than 8250 K for the BSS.

The typical log $g$ values for the WDs reported in the GCs (NGC 6397, NGC 6752, 47 Tuc), based on the spectra from earlier studies \citep{Moehler2000, Moehler2004, Knigge2008} lie in the range 7.5 - 7.8. The log $g$ values available in the Koester models that fall in this range are 7.5 and 7.75. We found that, for a fixed temperature and scaling factor, the SED fit is insensitive to the above two log $g$ values. This shows that the log $g$ value is not well constrained by the SED fit. We calculated the mass of the WD for two different log values (7.5 and 7.75) using the relation:

\begin{equation}
    \frac{M}{M_{\odot}}= \bigg(\frac{g}{g_{\odot}}\bigg)  \bigg(\frac{R} {R_{\odot}}\bigg)^{2}
\end{equation} 

For a fixed $T_{eff}$ and R of the WD given in Table \ref{sedpar}, we found that the mass of the WD varies from 0.5 - 0.9 $M_{\odot}$ for log $g$ values of 7.5 and 7.75. We assumed log $g$ = 7.5 in the SED fit as it corresponds to a WD mass $\sim$ 0.51 $M_{\odot}$ for the given radius, which is close to the average mass of WDs (0.53 $\pm$ 0.02 $M_{\odot}$) suggested in GCs \citep{Renzini1988, Renzini1996}. 

 \begin{table}
\centering
\caption{SED fit parameters of NH 84}
\setlength{\tabcolsep}{8pt}
\begin{tabular}{ccc}
\hline
Parameters & BSS & WD \\\hline
$T_{eff}$ &  8000 $\pm$ 250 K & 32000 $\pm$ 2000 K\\
log $g$ & 4.0 $\pm$ 0.5 & 7.5 - 7.75\\
$M_{d}$ & 4.1E-24 & 8.8E-28\\\hline
%$L/L_{\odot}$  & 7.58 $\pm$ 1.10 & 0.42 $\pm$ 0.11	\\
%$R/R_{\odot}$&	1.44 $\pm$ 0.05 & 0.021 $\pm$ 0.001\\\hline
\label{sedpar}
\end{tabular}
\end{table}

\section{Discussion}
\label{discus}
From the SED analysis, we found that 6 out of 14 BSSs show significant excess in UV, among which one is a known contact binary (NH 19) and one is a SX Phe variable (NH 49). We studied the UV excess of BSS NH 84 in detail, as we have the radial velocity measurements from Gemini spectra in addition to PM from GAIA DR2. The rest of the UV excess BSSs will be studied in detail in future.  

The mass of the BSS NH 84 is $\sim$ 1.1 M$_{\odot}$ when compared with the Padova isochrone (Figure \ref{cmd}). The mass of WD ranges from 0.5-0.9 M$_{\odot}$ as described in Section \ref{error}. This suggests that the hot component of the BSS NH 84 is more likely to be a C/O WD as inferred from the fit parameters and the associated uncertainties. 

%We also inferred the possible mass of the WD using different C/O and He WD models. 
A comparison of the L and $T_{eff}$ of the hot companion of BSS NH 84 with the Bergeron WD cooling models which are basically for C/O WDs \citep{Tremblay2011} suggests that the mass of the WD varies from 0.45-0.62 M$_{\odot}$ with a cooling age $\sim$ 15 Myr. This indicates that the system might have recently undergone a MT. %We also compared it with the He WD models of very low metallicity Z=0.0002 given by \cite{Istrate2016} which are generated for Low Mass X-ray Binary (LMXB) type systems with a Neutron Star and WD in binary. If we consider these models, we get a WD mass of $\sim$ 0.44 M$_{\odot}$.
According to the initial-final mass relationship by \citep{Althaus2015} for low metallicity systems (Z=0.0001), the progenitor mass corresponds to 0.8 M$_{\odot}$ for a final WD mass of $\sim$ 0.51 M$_{\odot}$ (log $g$ = 7.5). This suggests that the progenitor mass is likely to be only slightly higher than the MSTO mass of the cluster. Thus, we speculate that the BSS could have formed as a result of a Case B or C MT \citep{Paczy1971}.
%The initial-final mass relation for WD evolutionary sequences given by \cite{Althaus2015} for low metallicity progenitors is available only till a final WD mass of $\sim$ 0.5 M$_{\odot}$. {\bf Thus, as per the above models, the WD mass is $<$ 0.5 M$_{\odot}$. In this case, it would require a progenitor with mass less than the MSTO mass, which will take more than the cluster age to evolve. It is possible that the binary evolution in this system has produced a lower mass WD, due to mass transferred to the BSS early in its evolution.
 The WD parameters obtained for BSS NH 84 are similar to the parameters derived by \cite{Knigge2008} for a BSS-WD system in 47 Tuc. This is the second BSS-WD system to be detected in a GC after the first detection of one such system in 47 Tuc \citep{Knigge2008}. 

We checked the PM membership of all the BSSs available in the catalog given by \cite{Sandquist} using GAIA DR2 and found 8 of them to be non-members. Of the 8 non-members, 3 of them (NH 64, 83, 86) are classified as quasars by \cite{sdss2012, quasar2015}. 3 BSSs (NH 85, 87, 89) which do not have PM information from GAIA DR2 are classified as galaxies by SDSS survey \citep{sdss2015}. These 6 sources (galaxies and quasars) are mainly located outside 2$r_{h}$ of the cluster. Thus, we find that $\sim$ 15\% of the BSS population reside outside 1$r_{h}$ corresponding to 12 sources. In this study, where we have identified 14 BSSs in FUV, $\sim$ 14\% (2 BSSs) of the BSS population lie outside 1$r_{h}$ of the cluster. This shows that the distribution of FUV detected BSSs is consistent with the distribution of optically identified BSSs in the cluster.

\cite{Beccari} found the radial distribution of BSSs in NGC 5466 to be bimodal with a centrally-concentrated and an outer BSS sub-population, with a minimum in the radial surface density distribution at about $r\approx180\arcsec$. They estimated the binary fraction in the cluster outskirts ($400\arcsec < r < 800\arcsec$) to be $\sim$ 5$\%$. NH 84 is located at $r\approx730\arcsec$ ($\sim 8.5 r_{c}$, \cite{Mclaugh2005}) from the cluster center which is at half the distance of the tidal radius of the cluster ($r_{t} \sim 1580\arcsec$ \citep{Miocchi2013}). They concluded that the unperturbed evolution of primordial binaries could be the dominant formation mechanism for the BSSs in the low density environments. According to \cite{Ferraro2012}, NGC 5466 is in its dynamical infancy. The binaries present in the cluster outskirts has just recently begun to segregate towards the GC center. In light of the bimodal radial BSS density distribution, we speculate that NH 84 might be a MT binary system that has not yet experienced any significant dynamical interaction with the ambient stellar population and has evolved in relative isolation so far. The consistent picture of the location of NH 84 in NGC 5466 and its dynamical age together with the radially bimodal density distribution suggests that MT is one of the primary BSS formation mechanism in the low density environments \citep{Knigge2009, Geller2011, Leigh2013, Gosnell2014}.

\section{Summary and Conclusions}
\label{sum}
The first results for the metal-poor globular cluster NGC 5466 from UVIT are presented here. The results are based on our observations in four filters of UVIT (2 FUV and 2 NUV) along with Gemini spectra. 

Our study has led us to the following conclusions:

1. We detected 14 BSSs in NGC 5466, all of which have measured fluxes in all four UVIT filters and are likely proper motion members according to GAIA DR2.

2. The parameters of the BSS NH 84 obtained from the GMOS-N spectrum are T$_{eff}=$ 8000$^{+1000}_{-250}$K and log $g=$  4.0 $\pm$ 0.5. It is a radial velocity ($\sim$ 128 $\pm$ 30 km/s) member. 

3. The SED decomposition analysis found the presence of a hot component in the SED of BSS NH 84. The hot component is found to have a temperature of T$_{eff}=$ 32000 $\pm$ 2000 K and a radius $\sim$ 0.02 R$_{\odot}$ suggesting it to be a WD. 

4. NH 84 is the first BSS-WD candidate found in the outskirts of a low density GC. This is the second BSS-WD system reported in a GC. As NGC 5466 is a dynamically young cluster, this result suggests a MT pathway for BSS formation in low density environments.

\section{Acknowledgements}
We thank the anonymous referee for a very insightful and helpful report which improved the paper. We thank Avinash Singh for helping in python programming. This publication uses the data from the AstroSat mission of the Indian Space Research Organisation (ISRO), archived at the Indian Space Science Data Centre (ISSDC) which is a result of collaboration between IIA, Bengaluru, IUCAA, Pune, TIFR, Mumbai, several centres of ISRO, and CSA. Additionally, this research is based on the observations obtained at the Gemini Observatory, which is operated by the Association of Universities for Research in Astronomy, Inc., under a cooperative agreement with the NSF on behalf of the Gemini partnership: the National Science Foundation (United States), the National Research Council (Canada), CONICYT (Chile), Ministerio de Ciencia, Tecnolog\'{i}a e Innovaci\'{o}n Productiva (Argentina), and Minist\'{e}rio da Ci\^{e}ncia, Tecnologia e Inova\c{c}\~{a}o (Brazil). This publication makes use of VOSA, developed under the Spanish Virtual Observatory project supported by the Spanish MINECO through grant AyA2017-84089. This research also made use of Topcat, Matplotlib, IPython, and Astropy, a community-developed core Python package for Astronomy. 

\bibliographystyle{aasjournal}
\bibliography{ngc5466_ref.bib}

\end{document}